\documentclass[final,5p]{elsarticle} 
\usepackage{graphicx}
\usepackage{epsf}       
\usepackage{epsfig}       
\usepackage{amssymb}
\usepackage{rotating}
\usepackage{amsmath}
\usepackage{color}
\usepackage{siunitx}
\usepackage{lineno}
\usepackage{enumitem}
\usepackage{booktabs}
\usepackage{hyperref}
\usepackage[tableposition=top]{caption}
\usepackage[super]{nth}
\usepackage{lipsum}
\usepackage{mathtools}
\usepackage{cuted}

\newcommand{\acot}{\mathrm{acot}}
\newcommand{\cosec}{\mathrm{cosec}}

\newcommand{\bs}{\boldsymbol}
\newcommand{\diag}{\mathop{}\!\mathrm{\bf diag}}
\newcommand{\eqex}{\stackrel{!}{=}}

\newcommand*{\diff}{\mathop{}\!\mathrm{d}}
\newcommand{\GamTheta}{{\Gamma_{\! \theta\theta}}}
\newcommand{\GamPhi}{{\Gamma_{\! \phi\phi}}}
\newcommand{\GamThPhi}{{\Gamma_{\! \theta\phi}}}
\newcommand{\deltak}{{{\vec \delta}_k}}

\newcommand{\hThetak}{{{\vec h}_{\theta_k}}}
\newcommand{\hPhik}{{{\vec h}_{\phi_k}}}

\newcommand{\Krho}{{K_{\!\rho}}}

\newcommand{\GTTF}{{GTTF}}
\newcommand{\Kalman}{{K\'alm\'an}}

\newcommand{\curv}{{\kappa}}
\newcommand{\curvmin}{{\curv_\textrm{min}}}
\newcommand{\curvhit}{{\curv_\textrm{hit}}}
\newcommand{\curvMS}{{\curv_\textrm{MS}}}
\newcommand{\curvMSreg}{{\curv_\textrm{MSreg}}}
\newcommand{\curvMSj}{{\curv_{\textrm{MS},j}}}
\newcommand{\curvloc}{{\curv_\textrm{loc}}}
\newcommand{\curvref}{\curv_\textrm{ref}}
\newcommand{\curvSa}{{\curv_\textrm{01}}}
\newcommand{\curvSb}{{\curv_\textrm{12}}}
\newcommand{\curvloss}{{\curv_\textrm{loss}}}

\newcommand{\E}{\mbox{I\negthinspace E}}

\newcommand{\Covvb}{\vec  {\textrm{\bf C}}\textrm{\bf ov}}
\newcommand{\Covvv}{\vec {\vec {\textrm{C}}}\textrm{ov}}
\newcommand{\Covvvb}{\vec {\vec  {\textrm{\bf C}}}\textrm{\bf ov}}

\newcommand{\Covhh}{\bar {\bar  {\textrm{C}}}\textrm{ov}}
\newcommand{\Qvv}{\vec {\vec {Q}}}
\newcommand{\Vvv}{\vec {\vec {V}}}
\newcommand{\Qvvb}{\vec {\vec {\bs Q}}}
\newcommand{\Dvvb}{\vec {\vec {\bs D}}}

\DeclareMathOperator{\Var}{Var}
\DeclareMathOperator{\Cov}{Cov}

\usepackage{enumitem, kantlipsum}
\DeclareSIUnit\barn{b}

\DeclareFontFamily{OMX}{yhex}{}
\DeclareFontShape{OMX}{yhex}{m}{n}{<->yhcmex10}{}
\DeclareSymbolFont{yhlargesymbols}{OMX}{yhex}{m}{n}
\DeclareMathAccent{\wideparen}{\mathord}{yhlargesymbols}{"F3}

\usepackage{xcolor}

\usepackage{fixme}
\fxsetup{
    status=draft,
    author=,
    layout=inline,
    theme=color
}

\definecolor{fxnote}{rgb}{0.8000,0.0000,0.0000}
\colorlet{fxnotebg}{yellow}

\makeatletter
\renewcommand*\FXLayoutInline[3]{%
  \@fxdocolon {#3}{%
    \@fxuseface {inline}%
    \colorbox{fx#1bg}{\color {fx#1}\ignorespaces #3\@fxcolon #2}}}
\makeatother

\usepackage{apptools}
\AtAppendix{\renewcommand\appendixname{\relax}}

\title{A General Track Fit based on Triplets}

\author[unihd]{A.~Sch\"oning}

\ead{schoning@physi.uni-heidelberg.de}

\address[unihd]{Physics Institute, Heidelberg University, Im Neuenheimer Feld 226, 69120 Heidelberg, Germany
}

\journal{Nuclear Instruments and Methods A}

\begin{document}

{
  \begin{abstract}
  This paper presents  a general three-dimensional track fit based on hit triplets.
  The general track fit considers spatial hit and multiple Coulomb scattering uncertainties, and can also be extended to include energy losses. 
  Input to the fit are detector-specific triplet parameters, which contain information about
  the triplet geometry (hit positions),
  the radiation length of the material and the magnetic field.
  Since the solution is given by an analytical closed-form, it is possible to use the same fitting code for all kind of tracking detectors.

  Fitting formulas are given for the global track fit as well as for the local hit triplets.
  The latter allows filtering out triplets with poor fit quality at an early stage of track reconstruction.
  The construction and fit of local triplets is fully parallelisable, enabling accelerated computation with parallel hardware architectures.
  Formulas for the detector-specific triplet parameters are derived for the two most commonly used field configuration for tracking detectors, namely a uniform solenoidal field and gap spectrometer dipole.
  An algorithm to calculate the triplet parameters for an arbitrary magnetic field configuration is presented too.
  
  This paper also includes a discussion of inherent track fit biases.
  Furthermore, a new method is proposed to accelerate track fitting by classifying tracking regimes and using optimal fit formulas.
\end{abstract}

\maketitle

\section{Introduction} \label{sec:intro}
In nuclear and particle physics experiments, the precise determination of the track parameters for measuring charged particles is crucial.
Therefore an accurate tracking model  is required that takes into account all error sources of the measurement,
the most relevant being hit position errors, multiple Coulomb scattering (MS), energy losses and magnetic field errors.
What makes track fits so challenging is the fact that particles in the magnetic fieldd propagate along complex trajectories, which are highly non-linear.
For track reconstruction, the fit quality is the most important estimator for finding the correct hit combinations.
Especially in high-rate experiments, track finding is a major challenge due to large hit combinatorics.

The most commonly used track fit today is the \Kalman\ filter (KF)\cite{Kalman:1960:nal,Fruhwirth:1987fm}.
It uses a state vector to parameterize the track, which is updated with each measurement (hit), together with the quality of the track fit.
An advantage of the KF is its high flexibility, which allows for a wide range of applications.
Nowadays, many experiments employ an extended version of the Kalman Filter (KF) for track reconstruction, known as the combinatorial Kalman filter (CKF), which aims to identify the optimal hit combinations.

The KF, however, also has disadvantages:
the algorithm is recursive, and consequently not well suited for parallel computing.
This presents a significant challenge for accelerating track reconstruction using modern highly parallel computing hardware.
In addition, the KF does not provide the full covariance matrix of all hit positions, complicating its use for track-based detector alignment.
Considering that detector resolutions continue to improve with new tracking detector technologies, the software alignment of the detector system becomes increasingly relevant to fully exploit the potential of detectors.

For detector alignment, the General Broken Line (GBL) fit \cite{Blobel:2006yi,Kleinwort:2012np} is better suited as it inherently provides the full hit covariance matrix, which is required to determine correlations and allows for the identification of so-called \textsl{weak modes}. An example is the Millepede~II software tool \cite{Blobel:2006yh,Blobel:2011az}.
The basic concept of the GBL is to linearize an approximate solution and perform the track fit in a local (curvilinear) coordinate system defined by a reference (seed) solution.
MS as well as energy losses then show up as kinks in the transformed trajectory.
These kinks are minimized along with the hit residuals in the fit.
As the GBL is seeded and requires an approximate solution as starting point, it cannot be used for track finding.

The MS triplet fit \cite{ref:MSPaper} is an alternative track fit that uses a linearization approach quite similar to the GBL but does not require any seed or approximate solution.
Triplets of hits have the advantage that the reference trajectory for the linearization can be easily calculated from the triplet geometry itself, for example in a uniform magnetic field.
Furthermore, hit triplets are over-con\-strained, allowing the calculation of a triplet quality that can be used to reject fake hit combinations at an early stage of track reconstruction.
With single triplets, even a full track reconstruction in a high track multiplicity environment like FCC-hh is possible, as demonstrated in~\cite{Kar:2024wbi}.

Because the result of a single triplet fit can be written as a simple function of triplet-specific parameters, and the global track parameters can be calculated from simple sums of local triplet fits, the MS triplet fit is much faster than any other track fit.
Its parallelization capability makes the MS triplet fit ideal for parallel computing, for example on graphics processing units (GPUs).
However, since hit position errors are not included, the MS triplet fit is restricted to low-momentum tracks, where MS errors are dominant.
The MS triplet fit is used by the Mu3e experiment \cite{Mu3e:2020gyw}, which searches for the decay $\mu \rightarrow eee$ using muons decaying at rest.
Here, the triplet fit has been implemented for both offline reconstruction \cite{Kozlinskiy:2017wyl} and online track reconstruction on a GPU-based event filter~\cite{vomBruch:2017fqw}. \\

This work presents the General Triplet Track Fit (\GTTF), which is an extension of the MS triplet fit (\cite{ref:MSPaper}) and takes into account hit uncertainties as well as all correlations between different hit triplets.
Therefore, this work also goes beyond Ref.\cite{Schoning:2016cbj}, where hit uncertainties in the fit quality calculation were considered for individual triplets, but correlations between different triplets were neglected.
Interestingly, the solution of the GTTF can also be given in an analytical closed-form solution, similar to \cite{ref:MSPaper}.

A major difference between the GTTF and other track fits is the fit input. 
The \Kalman\
 filter and the GBL use hit positions as input.
In contrast, the \GTTF\ uses so-called \textit{triplet parameters} as input, which represent an interface to all kind of tracking detectors and provide a general description of the detector (triplet) geometry, including the hit position errors, the scattering material and the magnetic field.
For this reason, the \GTTF\ is universal as the same fitting code can be used for all tracking detectors and for all experiments.
Only the triplet parameters are experiment- and triplet-specific.

The most important advantage of \GTTF\ is the ability to perform triplet filtering during track reconstruction.
This, together with the ability to perform track fitting of triplets on a parallel computing architecture, offers great potential for accelerating track reconstruction in high particle rate experiments.
In addition, the \GTTF\ also provides the hit covariance matrix, making the fit ideal for track-based alignment.

Thanks to the analytical form of the result, the covariance can be directly calculated from the triplet geometries.
It is therefore relatively easy to calculate the tracking resolution for a given detector geometry, without the need for extensive simulation studies.
This feature greatly simplifies tracking detector design studies for future experiments.

With the triplet concept one can go even one step further;
from the triplet geometries simple \emph{tracking scale parameters} can be calculated, which can be used to define different \textit{tracking regimes}, for example \emph{MS dominated} and \emph{hit uncertainty dominated}.
Depending on the tracking regime, different numerical optimizations can be used to accelerate track fitting.
Furthermore, a tracking regime analysis can also help in identifying weaknesses of tracking detector designs.\\

The paper is organized as follows.
The fit methodology is introduced in \autoref{sec:method}.
The formulas for the global triplet track fit are derived in \autoref{sec:global_track_fit}, first for the general case, and then in the limit of dominant hit position errors and dominant MS errors.
Results for local triplet fits are given in \autoref{sec:local_fits}.
A detailed analysis of fitting biases as well as mitigation strategies, including a special regularized MS fit with reduced bias, is presented in \autoref{sec:Biases}.
The triplet parameters, which represent the input to the fit, are calculated in \autoref{sec:CalcTripletParameters} for the case of a uniform magnetic field.
Triplet parameter solutions for other setups (gap spectrometer dipole and other inhomogeneous magnetic fields) are discussed in the appendix.
A special solution obtained from MS in a zero magnetic field is als presented in \autoref{sec:CalcTripletParameters}.
Energy loss corrections and track fits including energy losses  are described in \autoref{sec:energy_loss}.
The potential for exploiting parallel computing for track fitting and track reconstruction using the triplet concept is presented in \autoref{sec:parallel},
and the tracking regime concept is introduced in \autoref{sec:regimes}.
Finally, \autoref{sec:summary} provides a summary.

\section{Fit Methodology and Triplet Representation}  \label{sec:method}
The track fit aims at fitting the total particle momentum, $p$, and the hit positions by simultaneously minimizing the MS angles and the hit position shifts.
The hit positions are given as shifts with respect to the  measured hit positions $\delta \vec{x}_k = \vec{x}_{\textrm{fit},k} - \vec{x}_{\textrm{meas},k}$, with $k$ being the hit index.
For a given magnetic field, this set of parameters ($p$ and all  $\delta \vec{x}_k$) contains the full information about the particle trajectory.

For track fitting, a $\chi^2$ function is defined that includes MS as well as spatial hit uncertainties according to:
\begin{align}
  \chi^2 & = \; \sum_{j=0}^{n_\textrm{scatt}-1}
               \frac{\theta_{\textrm{MS},j}^2}{\sigma_{\theta_\textrm{MS},j}^2} \ + \
  \sum_{j=0}^{n_\textrm{scat-1}} \frac{\phi_{\textrm{MS},j}^2}{\sigma_{\phi_\textrm{MS},j}^2} \nonumber \\
  & + \; 
  \sum_{k=0}^{n_\textrm{hit}-1} \
        \delta {\vec x}_k^{\;t} \, {\Vvv_k}^{-1} \, \delta {\vec x}_k .
\label{eq:global_chi2_general}
\end{align}
The first two sums\footnote{Throughout this work, counting of hits and triplets starts at 0.} run over all hit triplets (index $j$) and describe MS at the $n_\textrm{scatt}$ scattering points.
Throughout the work, it is assumed that the position of the scatterers agree with the position of the hits (detector layers).
Using spherical coordinates, the  MS kink is described by polar ($\theta_{\textrm{MS},j}$) and azimuthal ($\phi_{\textrm{MS},j}$) angles.
The projected MS angles are divided by the corresponding  expected errors ($\sigma_{\theta_\textrm{MS},j}$ and $\sigma_{\phi_\textrm{MS},j}$).
The third sum runs over all hits and describes the contribution from the hit position shifts (residuals).
For each hit, the hit position error is described by a $3 \times 3$ covariance matrix, $\Vvv_k$ (error ellipse\footnote{Throughout this work, single arrows (double arrows) denote vectors (matrices) in Euclidean space. Furthermore, $^t$ denotes a transposed vector in Euclidean space.}).

The kink angles in the MS terms of \autoref{eq:global_chi2_general} depend on the total particle momentum, as illustrated in \autoref{fig:kink_illustration}.
Instead of the particle momentum, $p$, the 3D curvature, defined by:
\begin{align}
\curv  & := \; \frac{q\, B}{p},
\end{align}
is used in the following\footnote{In many other papers, $\curv$ is used to denote the transverse curvature, which is henceforth denoted as $\curv_{\!\perp} = {q\, B}/{p_\perp} $ in this paper.}.
Note that for an inhomogeneous magnetic field, $B=B({\vec x)}$, $\kappa$ is position dependent even if total momentum is conserved.

Matter effects are  described by two parameters, an MS parameter and an energy loss parameter.
For each hit, both parameters are calculated from the effective path length  in the tracking layer material. 
Note that both matter effects have some momentum (energy) dependence.
The error of the MS angle, $\sigma_{\rm MS}$, depends on the momentum and velocity of the particle~\cite{ref:highland,Lynch:1990sq}.
Assuming that the particle is ultra-relativistic ($v \approx c$), MS is inverse proportional to the momentum
\begin{align}
\sigma_{\rm MS} \, & \propto \, \frac{1}{|p|}  
\label{eq:sigmams_c} ,
\end{align}
and one can define a \textit{MS parameter} according to:
\begin{align}
  b_\textrm{MS} \, & = \, \frac{\sigma_{\rm MS}}{|\curv|}  \, = \, \sigma_{\rm MS} \,  \left| \frac{p}{q \, B} \right|    \label{eq:def_b_ms} .
\end{align}

Similarly, an energy loss parameter, $\Delta E$, can be defined for each tracking layer (hit), 
accounting for the energy loss, for example due to ionization.
An advantage of the triplet fit is that both, the momentum  and the effective path length can already be derived from the triplet geometry
before fitting (see \autoref{sec:CalcTripletParameters}).

In the following, it is assumed that the total momentum of the particles is conserved.
Energy losses are included at a later stage and discussed in \autoref{sec:energy_loss}.

\subsection{Triplet Parameters} \label{sec:triplet_parameters}
\begin{figure}[tb]
   \begin{picture}(0.5\textwidth,130)
     \put(0,0){
       \epsfig{file=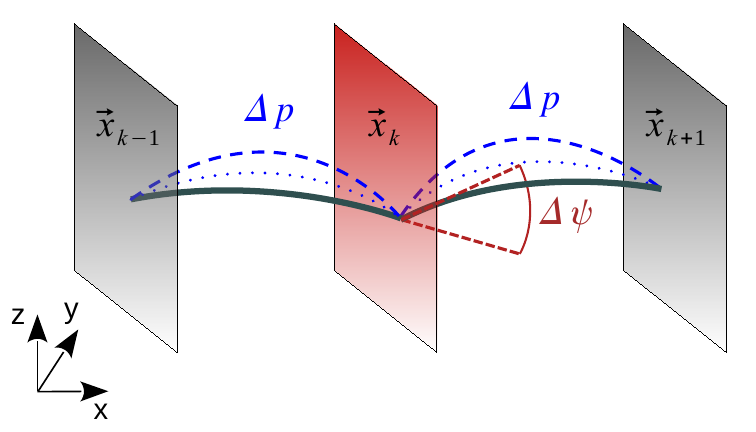,width=0.45\textwidth}
     }
   \end{picture}
   \caption{Sketch of the curvature (momentum) dependence of the kink angle in a triplet.
     The middle plane is the scattering plane in the triplet defined by the hits $\{k-1,\,k,\,k+1 \}$.
     The dashed and dotted trajectories show the momnentum varations of the solid trajectory. 
     The kink angle, indicated at the middle layer, has two projections  $\Delta {\psi} =( \Delta \theta$, $\Delta \phi)$, which are not shown.} 
\label{fig:kink_illustration}
\end{figure}
In a magnetic field, the trajectory between two consecutive hits is fully defined by the value of the total momentum\footnote{Note that there might be no solution for low total momentum tracks, and more than one solution for high momentum tracks, depending on the field configuration.} .
Consequently, the total momentum defines the kink angle $\Delta {\psi} =( \Delta \theta$, $\Delta \phi$) for a hit triplet, as shown in \autoref{fig:kink_illustration}.
Throughout this paper, a right-handed coordinate system is used, with the polar angle, $\theta$, being defined with respect to the $z$-axis, and the azimuthal angle, $\phi$, being defined with respect to the $x$-axis.
The two projections of the kink angle at detector layer $k$ are then defined as:
\begin{align}
  \Delta \theta(p) \; &= \; \Delta \theta(\curv) ~ := ~ \theta_{k,k+1} - \theta_{k,k-1} , \\
  \Delta \phi(p) \; &= \; \Delta \phi(\curv) ~ := ~ \phi_{k,k+1} - \phi_{k,k-1} ,
\end{align}
where the subscript ``${k,k-1}$'' (``${k,k+1}$'') indicates the particle direction\footnote{
  Note that in uniform magnetic fields the relation $\theta_{k,k'}=\theta_{k',k}$ holds for $k'=k\pm 1$, as the polar angle is an invariant.
  In contrast, the inequality $\phi_{k,k'} \ne \phi_{k',k}$ generally holds, due to the bending in the magnetic field.
   }
  at the detector plane before (after) the scattering at layer~$k$.
Both kink angles are functions of the momentum ($\cong$ 3D curvature), and for typical tracking detectors, these functions are transcendental.
This is, for example, the case for tracking in a uniform magnetic field or a spectrometer setup.
Both cases are discussed in \autoref{sec:CalcTripletParameters}.

\begin{figure}[tb]
   \begin{picture}(0.5\textwidth,145)
     \put(0,0){
       \epsfig{file=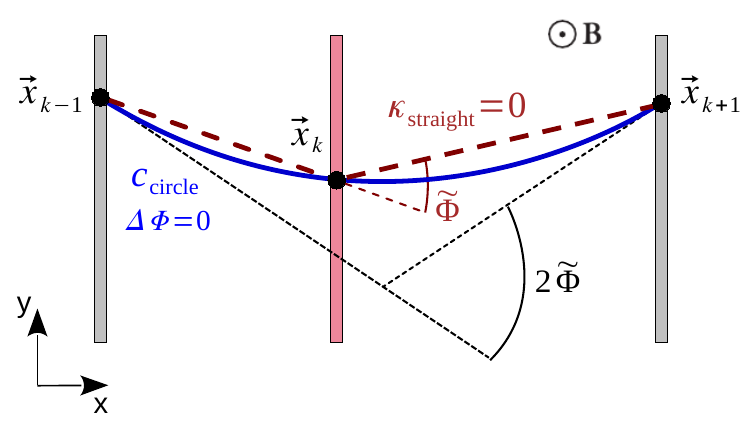,width=0.48\textwidth}
     }
   \end{picture}
   \caption{Sketch of a hit triplet in the bending plane of a uniform magnetic field. The \textsl{blue solid line} shows the solution for $\Delta \phi=0$ (zero kink angle); the \textsl{brown dashed line} shows the zero curvature solution ($\curv=0$).
     The triplet parameter $\tilde \Phi$ corresponds to the kink angle of the zero curvature solution, and is related  to the bending angle of the zero kink angle solution via $\Phi_{(\Delta \phi =0)} = 2 \, \tilde \Phi$. } 
\label{fig:triplet_parameters_geometry}
\end{figure}

A method for solving the non-linear functions $\Delta \theta(\curv)$ and $\Delta \phi(\curv)$, is to perform a linearization around a known solution.
Throughout the paper, the solution
\begin{align}
  \Delta \phi_\textrm{ref} \, := \,  \Delta \phi(\curv_\textrm{ref})=0 \label{eq:noMS}
\end{align}
  is used as \textit{reference} trajectory, corresponding to no MS in the $x$-$y$ plane, which is defined to be the \textit{main} bending plane.
The reference solution is described by the 3D curvature, $\curv_\textrm{ref}$, and has a non-vanishing polar kink angle  $\Delta \theta_\textrm{ref}:= \Delta \theta(\curv_\textrm{ref})$, in general.
The first order linearization around the reference solution then reads:
\begin{align}
  \Delta \phi \; & = \, \ \ \ 0 \ + (\curv-\curv_\textrm{ref}) \, \rho_\phi + O(\curv^2)
                               \;  \approx \; \tilde \Phi    \, + \,  \rho_\phi \,  \curv , \label{eq:linearization_phi} \\
  \Delta \theta \; &  = \; \Delta \theta_\textrm{ref} + (\curv-\curv_\textrm{ref}) \, \rho_\theta + O(\curv^2)
                               \;    \approx \;  \tilde \Theta  \, + \,  \rho_\theta \,  \curv \,  \label{eq:linearization_theta} .
\end{align}
The same ansatz was already used in Ref.\cite{ref:MSPaper}, with the only difference that the linearization was done as function of the 3D radius $R_\textrm{3D} = \curv^{-1}$, leading to marginally different numerical results for MS fits. 

The four linearization parameters $\tilde \Phi$,  $\tilde \Theta$, $\rho_\phi$ and $\rho_\theta$ are \textit{fundamental} parameters, which describe the curvature dependence of the triplet kink angles.
In the small bending limit, $\curv \rightarrow 0$, the fundamental triplet parameters
$\tilde \Phi$ and  $\tilde \Theta$ can be interpreted as central angle of the hit triplet (see \autoref{fig:triplet_parameters_geometry} for $\tilde \Phi$).
The  parameter $\rho_\phi$  is always negative and its absolute value can be interpreted as effective arc lengths of the triplet, as will be shown in \autoref{sec:CalcTripletParameters}.
The parameter $\rho_\theta$ is a small correction factor and has no simple geometrical interpretation.

\subsection{Representation of Hit Position Errors} \label{sec:Representation}

In \autoref{eq:global_chi2_general}, the hit positions and their uncertainties are given in global coordinates.
However, using local detector coordinates is often simpler and more intuitive if it comes to hit position uncertainties.
Without loss of generality, a transformation into local hit coordinates:
\begin{align}
  {\vec x}_k  \rightarrow  {\vec x}^{\,\prime}_k = T_k({\vec x}_k) \label{eq:transformation}
\end{align}
is possible, where the position of each hit, $k$, is described by local bases ($\vec u_k ,\, \vec v_k ,\, \vec w_k$).
In case that the bases are orthogonal, local hit residuals are obtained from the transformation:
\begin{align}
    {\vec \delta}_k = \Qvv_k \, \Delta \vec x_{k} ,
\end{align}
with $\Delta \vec x_{k}=\vec x_{k}^\textrm{\,meas}-\vec x_{k}^\textrm{\,fit}$ being the hit residuals in global coordinates, and $\Qvv_k$ being an orthogonal (rotation) matrix\footnote{
A transformation into local detector coordinates is highly convenient when the directions of hit position errors are uncorrelated, which is normally the case.}. 
The corresponding covariance matrix in local coordinates is given by:
\begin{align}
  \Vvv^{\,\prime}_k   &= \, {\Qvv_k}^{t} \, {\Vvv_k} \; \Qvv_k .
\end{align}

\begin{figure}[tb]
   \begin{picture}(0.5\textwidth,130)
     \put(0,0){
       \epsfig{file=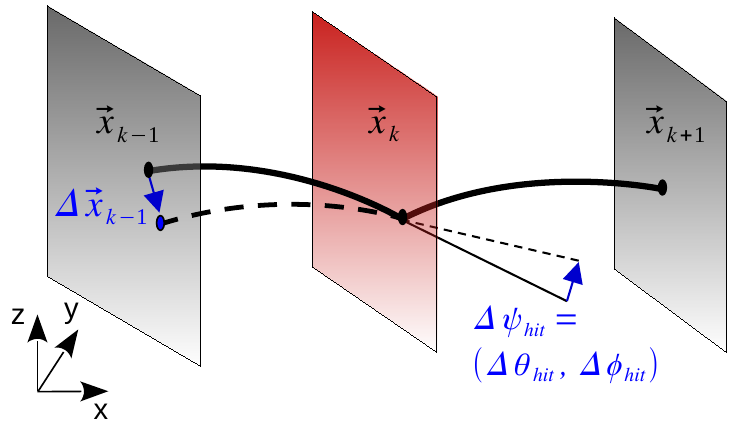,width=0.45\textwidth}
     }
   \end{picture}
   \caption{Illustration of the kink angle variation $\Delta \theta_\textrm{hit}$ and $\Delta \phi_\textrm{hit}$ at the scattering layer $k$ for a variation of the hit position in layer $k-1$.} 
\label{fig:kink_variation_hit}
\end{figure}

The variation of a hit position leads to a change of the kink angles as shown in \autoref{fig:kink_variation_hit}.
In practically all tracking devices, the hit position errors are significantly smaller than the distance between the hits (tracking layers), and the kink angles only weakly depend on the hit positions. 
The hit position-induced kinks can then be parameterized using the linearization ansatz:
\begin{align}
 \Delta   \theta_\textrm{hit} & = \; \sum_{k=0}^2   \, \hThetak \, \deltak  \label{eq:hitThetakink} \ ,
  \\
  \Delta   \phi_\textrm{hit} & = \;  \sum_{k=0}^2 \, \hPhik \, \deltak  \label{eq:hitPhikink} \ ,
\end{align}
with 
$\hPhik$ and $\hThetak$ being three vectors defined as directional gradients of the three hit positions:
\begin{align}
  \hThetak  &= \; {\vec \nabla}_{\deltak} \Delta \theta({\vec x}_{k})  \nonumber  \\
            &\approx \;  {\vec \nabla}_{\deltak} \tilde \Theta({\vec x}_{k}) \, +  \, \curv_\textrm{ref} \, {\vec \nabla}_{\deltak} \rho_\theta({\vec x}_{k})   \label{eq:hitGradientTheta} \ ,
  \\
  \hPhik  &= \;   {\vec \nabla}_{\deltak} \Delta \phi({\vec x}_{k})    \nonumber  \\
            &\approx \;  {\vec \nabla}_{\deltak} \tilde \Phi({\vec x}_{k}) \, +  \, \curv_\textrm{ref} \, {\vec \nabla}_{\deltak} \rho_\phi({\vec x}_{k}) \label{eq:hitGradientPhi}  \ .
\end{align}
The directional hit gradients can be determined numerically by shifting the hits by $1$-sigma of the hit position error into the three orthogonal directions and re-calculating the triplet parameters.

Finally, the MS angles of a triplet entering \autoref{eq:global_chi2_general} are expressed  as function of the four fundamental triplet parameters, the 3D curvature and the hit position-induced kinks:
\begin{align}
\theta_\textrm{MS} \;&=\;  \tilde \Theta  +  \rho_\theta \, \curv -  \Delta \theta_{\textrm{hit}}  \label{eq:MS_theta_linearization} \\
\phi_\textrm{MS} \;&=\;  \tilde \Phi  +  \rho_\phi \, \curv -  \Delta \phi_{\textrm{hit}} . \label{eq:MS_phi_linearization}
\end{align}

For the general fit of a single triplet, a total of 23 parameters are required.
These are the four fundamental triplet parameters, the $3 \times 3$ components of the hit gradients, the corresponding hit position errors, and one material parameter.
For MS fits where hit uncertainties are neglected, the number of parameters reduces to only five.

\section{Global Triplet Track Fit} \label{sec:global_track_fit}
Using the\ triplet parameters and the local hit coordinate representation introduced in the last section, 
the $\chi^2$ function (\autoref{eq:global_chi2_general}) can be re-written as:
\begin{align}
  \chi^2(\curv,\vec{\bs \delta})
  & = \; \sum_{j=0}^{n_\textrm{hit}-3} \frac{\left( \tilde \Theta_{j}  +  {\rho_\theta}_j \, \curv_j  
         -  \Delta \theta_{\textrm{hit},j}(\vec{\bs \delta}) \right)^2} {\sigma_{\theta_\textrm{MS},j}^2} \nonumber \\
  & + \;  \sum_{j=0}^{n_\textrm{hit}-3}
        \frac{\left( \tilde \Phi_{j}  +  {\rho_\phi}_j \, \curv_j
        -  \Delta \phi_{\textrm{hit},j}(\vec{\bs \delta}) \right)^2} {\sigma_{\phi_\textrm{MS},j}^2}   \nonumber \\
&  + \;  
  \sum_{k=0}^{n_\textrm{hit}-1} \  
     {\deltak}^{\, t}  \, {\Vvv_k^\prime}^{-1} \, {\deltak}  .
  \label{eq:chi2_global_start}
\end{align}
The fit parameters are the 3D curvature, $\curv$, and the residuals\footnote{Here (and in the following), bold symbols refer to vectors (lowercase variables) and matrices (uppercase variables), either in hit or triplet space. Furthermore, within this work two different transposition signs are used. The $^t$~operator acts on Euclidean space, whereas the $^\top$~operator acts on both triplet and hit space, also including all directions of the hit position errors.
} $\vec{\bs \delta}= ( \vec{\delta}_0,  \vec{\delta}_1, ... , \vec{\delta}_{n_\textrm{hit}-1} )^\top$.
  The 3D curvature, $\curv$ is here defined with respect to a reference magnetic field, $B_\text{ref}$, according to:
\begin{align}
  \frac{q}{p} \, & = \, \frac{\curv}{B_\text{ref}} \, = \, \frac{\curv_j}{B_j} ,
\end{align}
with $B_j$ being the local magnetic field strengths at triplet~$j$.
Note that for inhomogeneous magnetic fields, the field dependence of $\curv$ can be ``absorbed'' by the $\rho$ coefficients in \autoref{eq:chi2_global_start} by replacing: $\rho_j  \rightarrow \rho_j^\prime \, B_\textrm{ref}/B_j $.
In the following, it is assumed that such an replacement has been made.

In \autoref{eq:chi2_global_start}, the momentum dependence of the MS uncertainties ($\sigma_{\theta_\textrm{MS}}$ and $\sigma_{\phi_\textrm{MS}}$)
is deliberately neglected\footnote{Neglecting the momentum dependence  by setting $\sigma_\textrm{MS}$ constant, leads to a small momentum bias in track fits for MS-dominated particles, see also \autoref{sec:Biases}.}   in order to have at most quadratic terms as function of $\curv$ and $\vec{\bs \delta}$.
To bring \autoref{eq:chi2_global_start} in a more legible form, it is convenient to define two vectors, which contain all fundamental triplet parameters of a track:
\begin{align}
   \bs \rho  \;& = \; (\rho_{\theta_0} , \;  ... \; , \; \rho_{\theta_{n_\textrm{hit}-3}} ; \, \rho_{\phi_0} ,  ... , \; \rho_{\phi_{n_\textrm{hit}-3}})^\top , \nonumber \\
  \bs{\tilde \Psi} \;&= \; ({\tilde \Theta}_{0}, \;  ... \; , \; {\tilde \Theta}_{n_\textrm{hit}-3}; \,   {\tilde \Phi}_{0}  , \;  ... , \;  {\tilde \Phi}_{n_\textrm{hit}-3} )^\top ,  \nonumber
\end{align}
and whose length is twice the number of triplets.
Moreover, \textit{precision matrices} are defined for MS and hit position errors:
\begin{align}
  \bs D_\textrm{MS} &=\;  \diag \left(
  \frac{1} {\sigma_{\theta_\textrm{MS},0}^2}, \; ... \; , \;
  \frac{1} {\sigma_{\theta_\textrm{MS},{n_\textrm{hit}-3}}^2};  \right. \nonumber \\ 
&\;  \hspace{1.71cm}  \left.  \frac{1} {\sigma_{\phi_\textrm{MS},0}^2}, \;   ... \; , \,
  \frac{1} {\sigma_{\phi_\textrm{MS},{n_\textrm{hit}-3}}^2}
  \right)  \nonumber , \\
  \vec{\vec{\bs D}}_\textrm{hit} & =\; \diag  \left( {\Vvv^\prime_0}^{-1}, \; {\Vvv^\prime_1}^{-1} , \; ...\; ,  \; {\Vvv^\prime_{n_\textrm{hit}-1}}^{-1} \right) . \nonumber
\end{align}
$\bs D_\textrm{MS}$ and  $\vec{\vec{\bs D}}_\textrm{hit}$ are diagonal matrices, whose ranks are $2\, n_\textrm{triplet}$  and  $3\, n_\textrm{hits}$, respectively.
Elements in the MS precision matrix $\bs D_\textrm{MS}$ are partially related, since:
\begin{align}
\sigma_{\theta_\textrm{MS},j} \,&= \, \sigma_{\textrm{MS},j}  , \\
\sigma_{\phi_\textrm{MS},j} \,&= \, \sigma_{\textrm{MS},j} / \sin{\hat \theta_j} ,   \label{eq:sigma_phi_ms}
\end{align}
with $\sigma_{\textrm{MS},j}$ being the MS angular error (\autoref{eq:sigmams_c}) of the $j^\textrm{th}$ triplet and $\hat \theta_j$ being the corresponding estimated average polar angle in the MS process.
The $1/\sin{\hat \theta_j}$ factor in \autoref{eq:sigma_phi_ms} is  a geometrical factor originating from the chosen spherical coordinate representation.
Note that the $1/\sin{\hat \theta_j}$ term is treated as a constant, since it enters only  as a weighting factor to the fit and the corresponding propagated polar angle uncertainties are negligible.

\begin{figure}[tb!]
  \begin{center}
   \begin{picture}(300,140)
     \put(-5,0){
       \epsfig{file=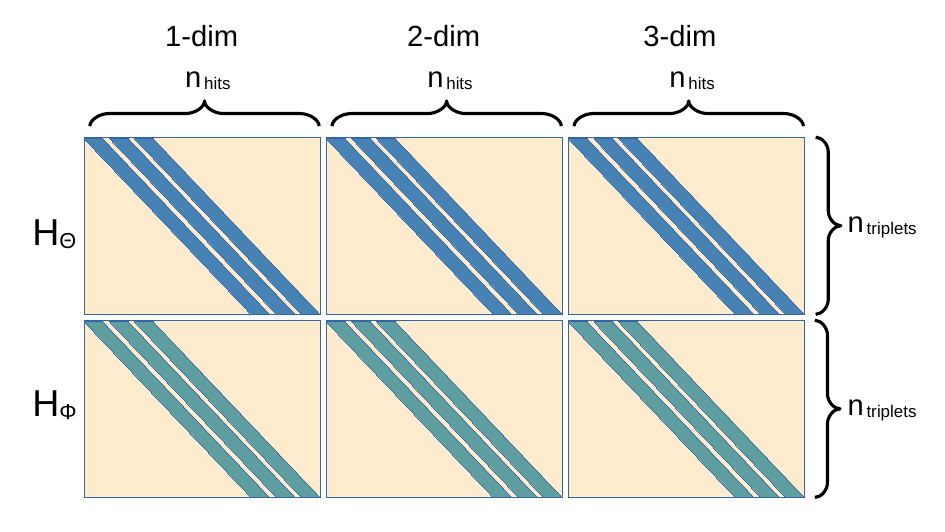,width=0.5\textwidth}
     }
   \end{picture}
   \caption{Sketch of the rectangular matrix  ${\vec {\bs H}}$.
     The horizontal block structure originates from the three directions of the hit position errors.
     The vertical block structure originates from the two projections of the MS angular error.
   Only elements in the blue and  turquoise bands, which originate from the three hits contributing to a triplet,  are non-zero.}
\label{fig:sketch_matrix_K}
  \end{center}
\end{figure}
To collect all hit gradients, defined by \autoref{eq:hitGradientTheta} and \autoref{eq:hitGradientPhi}, a hit gradient matrix (Jacobian) is defined, according to:
\begin{align}
  {\vec {\bs H}}   \;& :=\; \left(  {\vec {\bs h}}_{\theta}^{(0)} , \; ... \; , \; {\vec {\bs h}}_{\theta}^{(n_\textrm{hit}-3)} ; \right.
  \nonumber 
\\
        &\;    \hspace{2.61cm}  \left.  {\vec {\bs h}}_{\phi}^{(0)} , \; ... \; , \;  {\vec {\bs h}}_{\phi}^{(n_\textrm{hit}-3)} \right)^\top   \nonumber ,
\end{align}
where the vectors ${\vec {\bs h}}_{\theta}^{(j)}$ and ${\vec {\bs h}}_{\phi}^{(j)}$ are defined in hit-space and collect all hit gradients of triplet $j$:
\begin{align}
 {\vec {\bs h}}_{\theta}^{(j)}  \;& := \; \left( {\vec h}_{\theta_0}^{(j)} , \, {\vec h}_{\theta_{1}}^{(j)} , \, ... \, , \, {\vec h}_{\theta_{n_\textrm{hit}-1}}^{(j)} \right) \nonumber , \\
  {\vec {\bs h}}_{\phi}^{(j)}  \;& := \; \left( {\vec h}_{\phi_0}^{(j)} , \, {\vec h}_{\phi_{1}}^{(j)} , \, ... \, , \, {\vec h}_{\phi_{n_\textrm{hit}-1}}^{(j)} \right)  \nonumber .
\end{align}
Note that only elements with the indices $k=j,\, j+1,\,j+2$ (this are the hits forming the triplet)
are non-zero.
The matrix ${\vec {\bs H}}$ has in total $2 \, n_\textrm{triplet} \times 3 \, n_\textrm{hit}$ components; its structure is sketched in \autoref{fig:sketch_matrix_K}.

The $\chi^2$ function (\autoref{eq:chi2_global_start}) then reads in compact form:
\begin{align}
  \chi^2(\curv,\vec{\bs \delta}) \; & = \;
           \left( {\bs \Psi}  +  {\bs \rho} \, \curv  - \vec{\bs H} \vec{\bs \delta}    \right)^\top \!
         \bs D_\textrm{MS} \,
        \left( {\bs \Psi}  +  {\bs \rho} \, \curv  - \vec{\bs H} \vec{\bs \delta}    \right)
        \nonumber \\
&  \hspace{4cm}  + \;  
     \vec{ \bs {\delta}}^\top  \vec{\vec{\bs D}}_\textrm{hit}  \;  \vec{ \bs {\delta}}  .  \label{eq:chi2_global_compact_start}
\end{align}

Minimizing \autoref{eq:chi2_global_compact_start} results in a system of linear equations:
\begin{align}
  \left(
  \begin{array}{r}
  -  \bs \rho^\top \!   \bs D_\textrm{MS} \,  \vec{\bs \Psi}   \\
\vec{\bs H}^\top  \! \!  \bs D_\textrm{MS} \,  \vec{\bs \Psi}  
  \end{array} 
              \right)
   \ = \ \hspace{4.5cm}  &  \\
 \left(
  \begin{array}{rc}                                
    \bs \rho^\top \bs D_\textrm{MS} \, \bs \rho  &  - \bs \rho^\top \! \bs D_\textrm{MS} \, \vec{\bs H}    \\
   - \vec{\bs H}^\top \! \! \bs D_\textrm{MS} \,  \bs \rho  &   \vec{\vec{\bs D}}_\textrm{hit} +   \vec{\bs H}^\top \! \! \bs D_\textrm{MS} \, \vec{\bs H}  
  \end{array} 
              \right)
\
  \left(
  \begin{array}{c}                                
    \curv  \\
    \vec {\bs \delta} 
  \end{array} 
  \right)
  \begin{array}{c} \\ .
  \end{array} 
  \nonumber 
\end{align}

Solving the system above yields for the 3D curvature  and its variance:
\begin{align}
  \curvmin \;&=\;   - \frac{\bs \rho^\top  \bs K \bs \Psi} {\bs \rho^\top  \bs K \bs \rho} ,
                         \label{eq:global_fit_c3d}
  \\
  \sigma_\curvmin^2 \;& = \;   \frac{1} {\bs \rho^\top  \bs K \bs \rho} \label{eq:global_fit_sigma} ,
\end{align}
with $\bs K$ being the \textit{triplet precision matrix}.
Its inverse, the covariance matrix, is defined as:
\begin{align}
  \bs K^{-1} \;& =\;     \bs D_\textrm{MS}^{-1} \; + \; \vec{\bs H} \Dvvb_\textrm{hit}^{-1}   \vec{\bs H}^\top .  \label{eq:kmatrix_inverse}
\end{align}
$\bs K^{-1}$ combines the MS and hit position covariance matrices and is called \textit{triplet covariance matrix}.
Inversion of the triplet covariance matrix  is trivial in case of dominant MS errors since 
$\bs D_\textrm{MS}^{-1}$ is diagonal (see also \autoref{sec:global_fit_MS}).
The inversion is more involved if spatial hit uncertainties contribute.
Note that the matrix $\vec{\bs H} \Dvvb_\textrm{hit}^{-1}   \vec{\bs H}^\top$ has a $2 \times 2$ block structure with penta-diagonal sub-matrices.\footnote{
  The penta-diagonal structure of the sub-matrices can be exploited for large matrices where the computational effort for the inversion scales linearly with the number of hits (tracking layers).
}

The residuals and the corresponding covariance matrix are calculated as: 
\begin{align}
  {\vec {\bs \delta}}_\textrm{min}  \;&=\;    \Dvvb_\textrm{hit}^{-1} \vec{\bs H}^\top  \; \bs  \Krho \; \bs \Psi \label{eq:global_fit_delta}, \\
 \Covvvb_{\delta_\textrm{min}} \;& = \;  \Dvvb_\textrm{hit}^{-1}  - \; \Dvvb_\textrm{hit}^{-1}  \vec{\bs H}^\top \;  \bs  \Krho \; \vec{\bs H} \; \Dvvb_\textrm{hit}^{-1}  \label{eq:global_fit_cov},
\end{align}
with
\begin{align}
\bs  \Krho  \;&=\;  \left( \bs K   - \frac{\bs K \bs \rho \bs \rho^\top \bs K}   { \bs \rho^\top \bs K  \bs \rho}   \right)  .  \label{eq:global_Krho}
\end{align}
Note that the matrix $\bs  \Krho$ only exists for $\det(\bs K) \ne 0$.

Finally, the fit quality is given by:
\begin{align}
  \chi^2_\textrm{min} \;& =\;     \bs \Psi^\top \! \bs K \bs \Psi \; - \;  \frac{ (\bs \rho^\top \!  \bs K \bs \Psi)^2}{  \bs \rho^\top \!  \bs K \bs \rho  } \nonumber \\
   \;& =\; \bs \Psi^\top \! \bs \Krho \bs \Psi
  \label{eq:global_fit_chi2} .
\end{align}
The first term in the first line accounts for the kink angles of the infinite momentum solution whereas the second term describes the improvement of the fit quality by fitting the hit positions and the 3D curvature.
The second line of   \autoref{eq:global_fit_chi2} suggests that $\bs \Krho$ can be interpreted as post-fit precision matrix for the kink angles.

Note that with \autoref{eq:global_fit_c3d}, and \autoref{eq:global_fit_delta} the trajectory is fully determined from the first to the last hit.

\subsection{Global Fit for Dominant Hit Position Errors}
In the limit of dominating hit position errors, the MS errors can be neglected:
 $||\bs D_\textrm{MS}||^{-1} \rightarrow 0$.
 The solution looks very similar to the general case discussed above,
 and is given in \autoref{app:hit_position_errors}, for completeness.

\subsection{Global Fit for Dominant MS Errors}
\label{sec:global_fit_MS}
In the case of dominant MS errors, the hit position errors can be neglected.
By replacing the triplet covariance matrix by the MS covariance matrix, $\bs K \rightarrow \bs D_\textrm{MS}$, one obtains for the curvature and its variance:
\begin{align}
  \curvMS \;& =\;   -  \frac{\bs \rho^\top \bs D_\textrm{MS} \,  {\bs \Psi}}
                         {\bs \rho^\top \bs D_\textrm{MS} \,  {\bs \rho}}
                         \label{eq:global_MS_c3d} ,
\\
  \sigma_{\curvMS}^2 \;& =\;
                                    \frac{ 1 } {\bs \rho^\top \bs D_\textrm{MS} \,  {\bs \rho}}
        \label{eq:global_MS_sigma} ,
\end{align}
and for the fit quality:
\begin{align}
  \chi^2_\textrm{MS}  \;& = \;  {\bs \Psi}^\top  \bs D_\textrm{MS} \;  {\bs \Psi}
                                 \; - \; \frac{  ({\bs \rho}^\top  \bs D_\textrm{MS} \;  {\bs \Psi})^2  }{{\bs \rho}^\top  \bs D_\textrm{MS} \;  {\bs \rho}} .
        \label{eq:global_MS_chi2}
\end{align}
Due to the diagonal form of $\bs D_\textrm{MS}$, the global track curvature and the fit quality can be written as simple error-weighted sums of local triplet quantities:
\begin{align}
\curvMS  \;& = \; \sigma_{\curvMS}^2 \;  \sum_{j=0}^{n_\textrm{triplet}-1} \frac{\curvMSj^2}{\sigma_{\curvMSj}^2} , \\
\frac{1}{\sigma_{\curvMS}^2}  \;& = \;   \sum_{j=0}^{n_\textrm{triplet}-1} \frac{1}{\sigma_{\curvMSj}^2} , \\
  \chi^2_\textrm{MS}  \;& = \; \sum_{j=0}^{n_\textrm{triplet}-1} \chi^2_{\textrm{MS},j} \;
                          + \; \sum_{j=0}^{n_\textrm{triplet}-1} \frac{(\curvMS-\curvMSj)^2}{\sigma_{\curvMSj}^2}  , \label{eq:global_MS_chi2j}
\end{align}
where the indexed parameters denote the results obtained from the local triplet fits, which can be given in analytical closed-form (see \autoref{sec:local_fits}).
Note that the fit quality (\autoref{eq:global_MS_chi2j}) has two terms: a sum over the individual triplet qualities and a weighted sum over the curvature residuals (curvature consistency term\footnote{The combination of triplets using the MS fit was first discussed in Ref.\cite{ref:MSPaper} (Equation~39 therein) where, however, the curvature consistency term  is not given.}).

\subsubsection*{Fit Quality Relations}
For dominant MS errors, the following inequality can be derived from \autoref{eq:global_MS_chi2j}:
\begin{align}
\chi^2_\textrm{MS} & \ge \sum_{j=0}^{n_\textrm{triplet}-1} \chi^2_{\textrm{MS},j} .  \label{eq:MS_inequality}
\end{align}
By summing up the individual triplet fit qualities an lower limit on the global track fit quality is obtained.
\autoref{eq:MS_inequality} can therefore be used to reject bad track candidates already at triplet level, thus
 accelerating track reconstruction.

The fit quality of the MS fit is also related to the quality of the general fit:
\begin{align}
  \chi^2_\textrm{min} \;& =\;   \chi^2_\textrm{MS} \, + \, \frac{(\curvmin-\curvMS)^2}{\sigma_{\curvMS}}
                          \, - \,  \vec{\bs \delta}^\top_\textrm{min}  \vec{\vec{\bs P}}  \, \vec{\bs \delta}_\textrm{min} \label{eq:global_relation_general_MS} ,
\end{align}
with
\begin{align}
  \vec{\vec{\bs P}} \;&=\;  \vec{\bs H}^\top \!  {\bs D}_\textrm{MS} \, \vec{\bs H} + \Dvvb_\textrm{hit} 
\end{align}
being the \textit{adjoint triplet precision matrix}\footnote{
  The triplet precision matrix $\bs K$  and the adjoint triplet precision matrix $\bs P$ are related by:
  $\vec{\vec{\bs P}}^{-1} \vec{\bs H}^\top {\bs D}_\textrm{MS} = \Dvvb_\textrm{hit}^{-1}  \vec{\bs H}^\top \bs K$.
}, which has the rank $ 3 n_\textrm{hit}$.
As non-zero hit residuals can only result from a  $\chi^2$-improvement, the MS fit quality poses an upper limit for the general fit quality:
\begin{align}
  \chi^2_\textrm{min} \; & \le \;   \chi^2_\textrm{MS} \label{eq:chi2min_inequality}.
\end{align}
This relation is of high relevance for fast track finding:
Since the computational effort for the MS fit is significantly lower than for the general fit, which involves matrix inversion, it is often advantageous to perform the MS fit first, see discussion in \autoref{sec:regimes}.

\subsection{Large Hit Position Uncertainties}
Results obtained with GTTF linearization ansatz (\autoref{eq:linearization_phi} and \autoref{eq:linearization_theta} are only valid
if the hit position uncertainties are small compared to the distance of track layers, and so-called \textsl{rotational triplet uncertainties} can be neglected.
However, these rotational uncertainty can be significant for strip detectors.
Correction factors to include this uncertainty are given in \autoref{app:strip_detectors}.

\subsection{Track Parameters For Track Extrapolation}
The output of the GTTF are the curvature (momentum) and all hit residuals.
These parameters define the full trajectory, from the first to the last hit.
For track extrapolation or vertexing, also the track direction and the corresponding covariance matrix need to be known.
Both are position depend and, in general, complicated functions of the hit positions and the curvature.

For the special case of a uniform magnet field, the calculation of all track parameters and the corresponding covariance matrix is shown in \autoref{app:track_extrapol}.
The track parameter calculation is very similar for other magnetic field configurations.

\subsection{Additional Material in Tracking Volume}
Additional material, which affects the particle tracjectory by MS or energy loss, can be easily included in the GTTF by introducing so-called \textit{pseudo-hits} in \textit{pseudo-tracking} layers.
Pseudo-hits can be calculated by intersecting an already existing, approximate solution of the trajectory with the material layer, and by assigning (sufficiently) large hit position uncertainties.
It can be easily proven that the fit result remains stable as the hit uncertainties in the pseudo-tracking layers approach infinity.
  
\section{Local Triplet Fit} \label{sec:local_fits}
The local triplet fit represents the simplest solution of the GTTF, and is highly relevant for seeding track reconstruction and filtering.
The solution is readily obtained from Equations~\ref{eq:global_fit_c3d} to~\ref{eq:global_fit_chi2}.
For a triplet, the covariance matrix, $\bs K^{-1}$, reduces to a $2 \times 2$ matrix.
The elements of this local covariance matrix, defined as
\begin{align}
  \bs K^{-1}_\textrm{loc} \;& = \; \left(
    \begin{array}{cc}                                
      \GamTheta^\star & \GamThPhi \\
       \GamThPhi & \GamPhi^\star 
    \end{array}
  \right) 
  \label{eq:K_local} , 
\end{align}
are given by:
\begin{align}
  \GamTheta^\star & := \;   \GamTheta \; + \; \sigma_{\theta_{\textrm{MS},k}}^2 \nonumber  \\
                  & := \;  \sum_{k \, \in \, \textrm{triplet}}\;  {\hThetak}^t \Vvv_k^\prime \;   \hThetak  \; + \; \sigma_{\theta_{\textrm{MS},k}}^2 \ , \label{eq:gamtheta} \\
  \GamPhi^\star & := \;   \GamPhi \; + \; \sigma_{\phi_{\textrm{MS},k}}^2 \nonumber  \\
  & := \;  \sum_{k \, \in \, \textrm{triplet}}\;  {\hPhik}^t \Vvv_k^\prime \; \hPhik  \; + \; \sigma_{\phi_{\textrm{MS},k}}^2  \ , \label{eq:gamphi} \\
\GamThPhi & := \;  \sum_{k \, \in \, \textrm{triplet}}\;  {\hThetak}^t  \Vvv_k^\prime \;  \hPhik \ . \label{eq:gamthphi}
\end{align}
The solution of the local triplet fit is  a function of those $\Gamma$ parameters,
 and the 3D curvature and its variance are given by:
\begin{align}
  \curvloc \, & = \nonumber \\
 & \hspace{-0.5cm} - \frac{{\tilde \Theta} \, \rho_\theta \,
                             \GamPhi^\star \,+ \, {\tilde \Phi} \, \rho_\phi \, \GamTheta^\star \,  - \, \GamThPhi \, ({\tilde \Phi} \rho_\theta + {\tilde \Theta} \rho_\phi) }
                             {\rho_\theta^2 \, \GamPhi^\star  \,+\, \rho_\phi^2 \, \GamTheta^\star \,  - \, 2  \rho_\theta \, \rho_\phi \, \GamThPhi   } ,
                             \label{eq:local_c3d_general}
  \\
  \sigma_{{\curv_\textrm{loc}}}^2 \; & = \; { \frac
     {\GamTheta^\star \; \GamPhi^\star - \GamThPhi^2}
     {\rho_\theta^2 \, \GamPhi^\star  \; + \;  \rho_\phi^2 \GamTheta^\star \; - \; 2 \rho_\phi \, \rho_\theta \GamThPhi  }} .
\label{eq:local_sigma_general}
\end{align}
For the local fit quality, one obtains:
\begin{align}
  {\chi^2_\textrm{loc}}  \;& =\;  \frac{({\tilde \Theta} \rho_\phi  - {\tilde \Phi} \rho_\theta  )^2   }
        {\rho_\theta^2 \, \GamPhi^\star     \, + \, \rho_\phi^2 \, \GamTheta^\star \, - \, 2 \rho_\phi \, \rho_\theta \, \GamThPhi  }              .
  \label{eq:local_chi2_general}
\end{align}
Furthermore, the residual vector of hit $k$ is given by:
\begin{align}
 {\vec \delta}_{{k},\text{loc}}
\;&=\; \Vvv^\prime_k    \,
  \frac{(\rho_\theta  \, \hPhik  - \rho_\phi \,  \hThetak  )}
  {(\rho_\theta {\tilde \Phi} - \rho_\phi {\tilde \Theta})}  \,  {\chi^2_\textrm{loc}}  
       \label{eq:local_delta_general}   .
\end{align}
The covariance matrix of the residuals can be written as:
\begin{align}
  \Covvvb_{\bs \delta} \;
  & =  \; \Dvvb_\textrm{hit}^{-1}
    \,  - \, \frac{\vec{\bs \delta}_\textrm{loc}  \,  \vec{\bs \delta}^\top_\textrm{loc} } {\chi^2} .  \label{eq:global_fit_cov2}
\end{align}
The first term in \autoref{eq:global_fit_cov2} contains the pre-fit hit position errors, whereas the second term describes their improvement by the track fit.

\subsection*{Special Cases for Local Triplet Fits}
In the case of dominant hit position  errors, the following substitutions are applicable:
\begin{align}
\GamTheta^\star & \rightarrow \GamTheta , \\
\GamPhi^\star & \rightarrow \GamPhi .
\end{align}
In other words, one obtains the same fitting formulas as for the general triplet fitting by simply removing the $^\star$ from all equations.

For dominant MS errors, the following substitutions  are applicable:
\begin{align}
\GamTheta^\star & \rightarrow \sigma_{\theta_{MS}}^2 \\
  \GamPhi^\star & \rightarrow \sigma_{\phi_{MS}}^2 \\
  \GamThPhi & \rightarrow 0 ,
\end{align}
and the local triplet fit formulas (\autoref{eq:local_c3d_general} to \autoref{eq:local_chi2_general}) further simplify.
The hit residuals (\autoref{eq:local_delta_general}) vanish by definition and the 3D curvature (\autoref{eq:local_c3d_general} simplifies to:
\begin{align}
\curvMS \;  & = \; - \; \frac{\rho_\phi \; {\tilde \Phi} \; \sin^2{\hat \theta}  + \rho_\theta
  \; {\tilde \Theta}  }{\rho_\phi^2 \; \sin^2{\hat \theta}  + \rho_\theta^2} 
\label{eq:c3D_MS_solution} ,
\end{align}
Note that \autoref{eq:c3D_MS_solution}  is independent of the angular MS error, and therefore does not depend on the amount of material at the scattering layer.

\section{Momentum Bias in Track Fits} \label{sec:Biases}

  The tracking model used by the GTTF and other standard fits, such as the \Kalman\ Filter and General Broken Lines,  is the same; they all include hit position and MS uncertainties.
  The only difference is the way how the tracking model is implemented or how linearizations are performed.

  Due to the momentum dependence of the MS uncertainty, see \autoref{eq:sigmams_c}, 
  a fitting bias towards higher momenta (smaller curvatures) naturally arises in all track fits.
  The reason is that the \emph{estimated} MS angular errors  ($\sigma_{\theta_\mathrm{MS}}$ and $\sigma_{\phi_\textrm{MS}}$ in \autoref{eq:global_chi2_general}) are smaller for high momentum tracks than for low-momentum tracks.
  This bias is an inherent property of track fits that include MS uncertainties and where the particle momentum is a fit parameter.

  The analytical form of the GTTF solution facilitates the study of fitting biases. 
  Below, a detailed study of fitting biases is presented and a method for bias mitigation is proposed. 

\subsection{Curvature Pull Distribution}
One way to check the goodness of a fit is by studying the pull distributions of the fitted parameters.
The curvature pull for fit $i$ is defined as:
\begin{align}
\textrm{g}(\curv)_{i} \, & = \, 
                       \frac {
                       \curv_{i} - \curv^\textrm{true}
                       }{
                       \sigma_{\curv,i}
                       } .    \label{eq:def_pull}
\end{align}
For a fit with statistically correct estimates of $\curv$ and $\sigma_{\curv}$, the arithmetic mean of the pull distribution is expected to vanish, $\E[\textrm{g}{(\curv)}]=0$, and the variance is expected to be unity, $\Var[\textrm{g}{(\curv)}]=1$.
In the following, the pull distribution is studied for the Local Track Fit and the  Global Track Fit in a scenario where MS errors dominate.
  
\subsubsection{Curvature Pull in the  Local Track Fit} \label{sec:local_bias}
For a single triplet, the curvature pull can be calculated by smearing the particle momentum with a Gaussian distribution and recalculating the curvature error from the smeared quantity.
For small curvature errors, i.e., $\sigma_{\curv}^2 \ll \curv^2$, one obtains:
\begin{align}
  \E[\textrm{g}(\curv)] & \approx \; -\frac{|\sigma_{\curv_\textrm{MS}}|}{\curv} .  \label{eq:uncertainty_bias}
\end{align}
The expected curvature pull is shifted toward smaller curvatures (corresponding to higher momenta).
It is interesting to note that the curvature itself is correctly fitted, i.e., $\E[\curv] = \curv^\textrm{true}$.
Since the bias of the curvature pull distribution is caused by the \textit{a priori} unknown MS uncertainty,
this effect is henceforth referred to as \textit{MS normalization bias}.

\subsubsection{Curvature Pull  in Global Triplet  Track Fit} \label{sec:weighting_bias}
For dominant MS angular errors, the 3D curvature in the global fit  becomes a weighted sum of the local triplet fits (see \autoref{sec:global_fit_MS}).
Due to the MS normalization bias discussed in the previous section, high momentum triplets receive a higher weight in the fit than  low-momentum triplets.
This leads to the so-called \textit{weighting bias} of the curvature.

For demonstration, the weighting bias has been studied for a simple detector geometry with equidistant tracking layers.
Using a toy Monte Carlo, the following relation was empirically found for the weighting bias:
\begin{align}
 \E[\curv] - \curv^\textrm{true} &=\; - \left(2-\frac{2}{n_\textrm{triplet}} \right) \;  \frac{\sigma_{\curv_\textrm{MS}}^2}{\curv^\textrm{true}} .   \label{eq:weightingbias}
\end{align}
with $n_\textrm{triplet}$ being the number of triplets combined.
The weighting bias depends quadratically on the curvature error and is negative.

All together,  the pull distribution (\autoref{eq:def_pull}) is influenced by two effects: the weighting bias, which leads to a shift in the  curvature, and the MS normalization bias, which creates an asymmetry in the pull distribution.

\subsection{Bias Mitigation in the Global Triplet Fit}  \label{sec:mitigation_bias}
The size of the bias depends on the accuracy with which the MS uncertainties are known prior to the fit.
For the global fit, there are several options to calculate the MS angular errors that enter the MS precision matrix $\bs D_\mathrm{MS}$ (listed in order of increasing precision):
\begin{itemize}
\item from triplet parameter based estimates of the curvature (momentum),
  for example using the simple relation:  $\curv^\textrm{est} = \tilde \Phi / \rho_\phi$,
\item from locally fitted curvatures derived before the global fit,
  $\curv^\textrm{est} = \curvloc$ (\autoref{eq:local_c3d_general}),
\item by repeating (``updating'') the global fit (\autoref{eq:global_fit_c3d} or \autoref{eq:global_MS_c3d}), where the  MS angular errors are calculated from the curvature obtained in the first fit: $\curv^\textrm{est} = \curv_\textrm{global}(\textrm{\nth{1} fit})$\footnote{The updating of the GTTF corresponds to the \textsl{smoothing} step of the KF.}.
\end{itemize}
Since the global fit has a significantly higher precision than the local fits, the bias is significantly reduced by using the third method.
The downside is that the precision matrix $\bs K$ must be re-calculated (matrix inversion) after updating the MS angular errors, essentially doubling the computational effort.

Note that the same fitting bias also occurs in the KF.
However, due to the gradually increasing momentum precision, the bias in the KF in smaller compared to the GTTF without updating.
After the final smoothing step (KF) and update of the MS uncertainties (GTTF), the fitting biases of both methods are expected to be identical, as they use the same MS uncertainties as input for the track fit.

\subsection{Regularized MS Fit} \label{sec:regularized_fit}
The weighting bias of the curvature can be reduced by explicitly including the momentum dependence of the MS angular errors (\autoref{eq:sigmams_c}) in the fit.
If included, minimization of the general $\chi^2$ formula (\autoref{eq:chi2_global_start}) yields a system of equations that contains non-linear $ \curv^2 \cdot \delta_k$ terms, which is  difficult to solve.
However, since the bias is mainly relevant for the case of prevailing MS errors,
hit position errors are ignored in the following.
The MS precision matrix can then be replaced  by a new matrix that depends only on the MS parameters, defined in \autoref{eq:def_b_ms}:
\begin{align}
  \bs D_\textrm{MS} &\rightarrow \;  \frac{1}{\curv^2} \bs B_\textrm{MS} , \nonumber \quad \quad \textrm{with} \\
\bs B_\textrm{MS}   & := \;               \diag \left(
  \frac{1} {b_{\textrm{MS},0}^2}, \; ... \; , \;
  \frac{1} {b_{\textrm{MS},{n_\textrm{hit}-3}}^2};  \right. \nonumber \\ 
&  \hspace{2.71cm}  \left.  \frac{\sin^2{\hat \theta_0}} {b_{\textrm{MS},0}^2}, \;   ... \; , \,
  \frac{\sin^2{\hat \theta_{n_\textrm{hit}-3}}} {b_{\textrm{MS},{n_\textrm{hit}-3}}^2}
  \right)  \nonumber .
\end{align}
The fit quality after linearization then reads  in compact form:
\begin{align}
  \chi_\textrm{MSreg}^2(\curv) &= \:
           \frac{
           ( {\bs \Psi}  +  {\bs \rho} \, \curv  )^\top
        \, \bs B_\textrm{MS} \,
           ( {\bs \Psi}  +  {\bs \rho} \, \curv )
           }
           {\curv^2} 
             \label{eq:chi2_MS_compact_regularized} ,
\end{align}
and minimization of above equation yields:
\begin{align}
  \curvMSreg \;& =\;   -  \frac{\bs \Psi^\top \bs B_\textrm{MS} \,  {\bs \Psi}}
                       {\bs \rho^\top \bs B_\textrm{MS} \,  {\bs \Psi}}
                         \label{eq:global_MSreg_c3d} ,
\\
  \sigma_{\curvMSreg}^2 \;& =\; \frac{(\bs \Psi^\top \bs B_\textrm{MS} \,  {\bs \Psi})^{3}}
                       {(\bs \rho^\top \bs B_\textrm{MS} \,  {\bs \Psi})^4} .
        \label{eq:global_MSreg_sigma}
\end{align}
Finally, the fit quality is given by:
\begin{align}
  \chi^2_\textrm{MSreg}  \;& = \; {{\bs \rho}^\top  \bs B_\textrm{MS} \;  {\bs \rho}}
                          \; - \; \frac{  ({\bs \rho}^\top  \bs B_\textrm{MS} \;  {\bs \Psi})^2  }
                           {{\bs \Psi}^\top  \bs B_\textrm{MS} \;  {\bs \Psi}} .
        \label{eq:global_MSreg_chi2}
\end{align}
For completeness, formulas for the regularized local triplet fit are given in \autoref{sec:regularized_local}.

The regularized MS fit has the advantage that the curvature pull distribution is free of bias, see \autoref{sec:comparison_ms_fits}.
Another advantage is that the fit depends only on the MS parameters ($\bs B_\textrm{MS}$), and no longer on the estimated MS errors ($\bs D_\textrm{MS}$).
Therefore, there is no need to estimate the MS angular error before (global) fitting.

\subsection{Comparison of Regularized and Unregularized MS Fits}
\label{sec:comparison_ms_fits}

The biases of the regularized and unregularized MS fit are quantitatively compared using an example:
A relativistic particle with a momentum of \SI{300}{MeV/c} is simulated in a uniform magnetic field of \SI{3}{\tesla} using three tracking layers over a total distance of \SI{60}{mm}.
The material thickness of the middle tracking layer corresponds to $\SI{2}{\percent}\,X_0$.
Assuming that hit position errors are negligible, the corresponding relative curvature (momentum) resolution of this setup is about \SI{23}{\percent}.

\begin{figure}[tb]
  \begin{center}
   \begin{picture}(240,180)
     \put(0,0){
       \epsfig{file=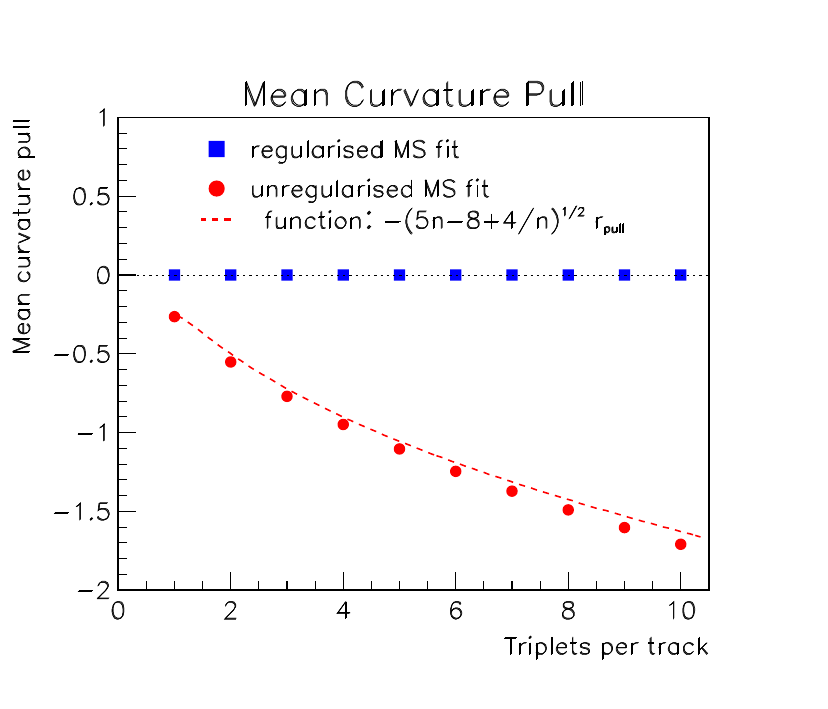,width=0.45\textwidth}
     }
   \end{picture}
   \caption{Simulated mean curvature pull (\autoref{eq:def_pull}) as function of the number of triplets  for a relative curvature (momentum) resolution of \SI{23}{\percent}.
     The results are shown for regularized (\textsl{blue squares}) and unregularized (\textsl{red points}) MS fits.
     The statistical errors are of the order of 1~per mil.
     The data points are compared with empirical functions.
     See text for more explanation.
     }
\label{fig:curvature_pull}
  \end{center}
\end{figure}

A comparison of the two fits reveals contrasting results:
For $n_\textrm{triplet}=1$, the unregularized MS fit has a mean curvature pull of $-25\,\%$ (towards higher momenta), see \autoref{fig:curvature_pull}, which further increases if more triplets are combined.
The regularized MS fit has no pull, as expected.
The negative curvature pull of the unregularized MS fit
is a combined effect of the MS normalization bias  (\autoref{sec:local_bias}) and the weighting bias (\autoref{sec:weighting_bias}), which can be described by a function that adds both sources quadratically, resulting in:
\begin{align}
 \E[g_{\curv}] &=\; - \, \sqrt{\frac{8n^2-5n+4}{n}} \;  \frac{\sigma_{\curv_\textrm{MS}}}{\curv^\textrm{true}} ,   \label{eq:reg_curvature_bias_many}
\end{align}
As shown in \autoref{fig:curvature_pull}, \autoref{eq:reg_curvature_bias_many} describes the simulation data reasonably well.

\begin{figure}[tb]
  \begin{center}
   \begin{picture}(240,180)
     \put(0,0){
       \epsfig{file=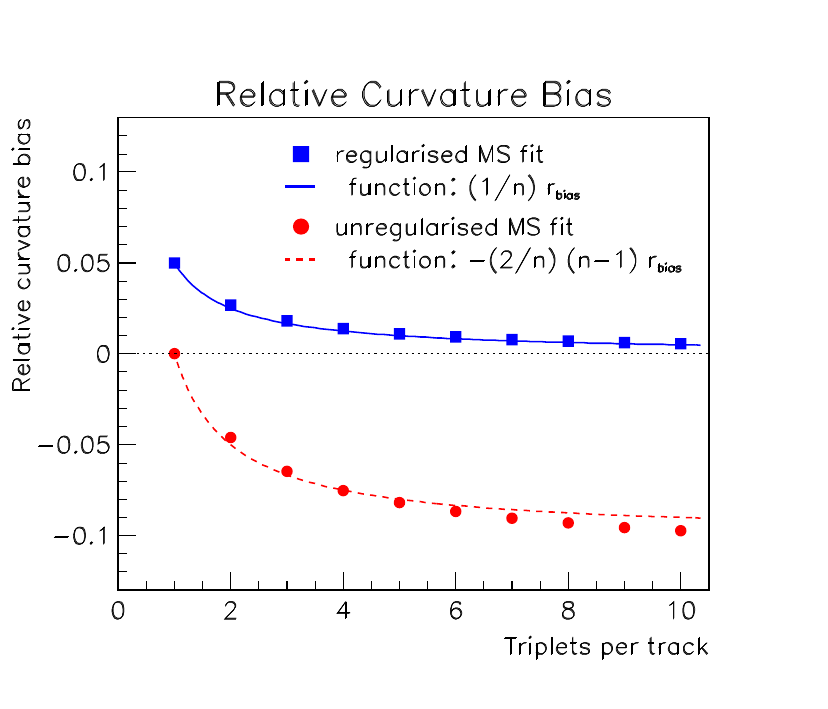,width=0.45\textwidth}
     }
   \end{picture}
   \caption{Simulated relative bias of the curvature,  $\E[\curvMSreg] / \curv^\textrm{true}-1$,   as a function of the number of triplets per track for a relative curvature (momentum) resolution of \SI{23}{\percent}.
     For more information, see caption of \autoref{fig:curvature_pull}.
   }
\label{fig:curvature_bias}
  \end{center}
\end{figure}

In \autoref{fig:curvature_bias}, the curvature bias of both fits is shown. 
For $n_\textrm{triplet}=1$, the unregularized MS fit has no curvature bias.
However, the regularized MS fit has a relative bias of $+5\,\%$ (towards lower  momenta).
This positive curvature bias compensates for the negative normalization bias,   so that the pull distribution of the regularized MS fit is free of errors.

Of particular interesting is the case when several triplets are combined ($n_\textrm{triplet}>1$).
For the unregularized MS fit, the weighting bias causes a shift towards negative curvatures (see \autoref{fig:curvature_bias}), which increases with $n_\textrm{triplet}$ and is quantitatively described by \autoref{eq:weightingbias}.
For the regularized MS fit, the curvature bias  decreases,  according to:.
\begin{align}
 \E[\curvMSreg]- \curv^\textrm{true} \, &=\, \frac{1}{n_\textrm{triplet}}  \, \frac{\sigma_{\curvMSreg}^2} {\curv^\textrm{true}} . \label{eq:bias_regularized}
\end{align}
Note that already for $n_\textrm{triplet}=2$, the (positive) bias of the regularized MS fit is smaller  than the  (negative) bias of the unregularized MS fit.

As demonstrated for the discussed example, both the MS normalization bias and the weighting bias can significantly deteriorate the curvature measurement if the unregularized MS fit is used. Therefore, preference should be given to the regularized fit if MS uncertainties dominate, since the unregularized MS fit requires an update to reduce the fitting bias.

\section{Calculation of Triplet Parameters} \label{sec:CalcTripletParameters}
The triplet parameters are the input for the global and local triplet fit (\autoref{sec:global_track_fit} and \autoref{sec:local_fits}); they are detector-specific and depend on the tracker geometry and the magnetic field.
In this section, the four fundamental triplet parameters ($\tilde \Phi$,  $\tilde \Theta$, $\rho_\phi$ and $\rho_\theta$) are calculated for the most commonly used detector design, which features tracking planes within a uniform solenoidal field (\autoref{sec:HomogeneousMagneticField}).
In section \autoref{sec:zero_field}, the case of zero magnetic field is discussed and formulas for measuring the particle momentum just through MS are given.

A gap spectrometer dipole as well as an algorithm for determining the triplet parameters for \textit{any} (inhomogeneous) magnetic field, are discussed in appendices (\autoref{sec:spectrometer_dipole} and \autoref{sec:GeneralFieldConfiguration}).

\subsection{Uniform Magnetic Field} \label{sec:HomogeneousMagneticField}

\begin{figure*}[tb]
   \begin{picture}(\textwidth,180)
     \put(0.125\textwidth,0){
       \epsfig{file=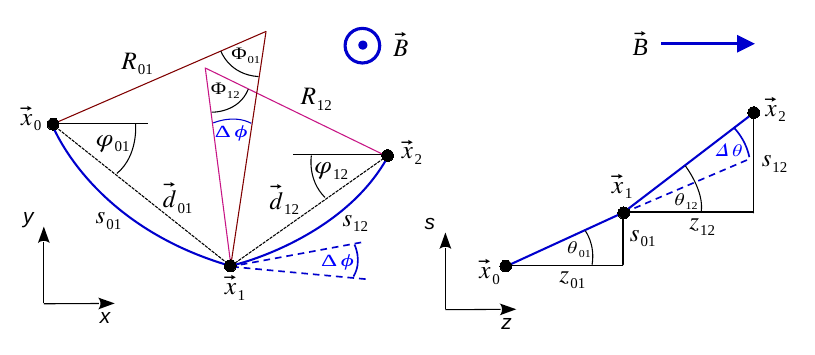,width=0.75\textwidth}
     }
   \end{picture}
   \caption{Hit triplet in a uniform magnetic field in the $x$-$y$ bending plane ({\it left}) and the
     $s$-$z$ non-bending plane ({\it right}).
     Hit positions are given by the three points ${\vec x}_0$, ${\vec x}_1$ and ${\vec
    x}_2$. $R_{01}$ and $R_{12}$ are the transverse bending radii before and after the middle layer. $s_{01}$ and $s_{12}$ denote  the transverse arc
  lengths and $\Phi_{01}$ and $\Phi_{12}$ the corresponding
  bending angles.  
  ${\vec d}_{01}$ and ${\vec d}_{12}$ denote the transverse distance vectors between hits in the transverse plane and $\varphi_{01}$ and $\varphi_{12}$ are the corresponding azimuthal angles.
  $\Delta \phi$ is the kink angle in the bending plane.
  In the non-bending plane, $z_{01}$ and $z_{12}$ define the longitudinal distances between
  adjacent hits, $\theta_{01}$ and $\theta_{12}$ the corresponding polar angles and
  $\Delta \theta$ is the kink angle. Modified from Ref.\cite{ref:MSPaper}.
}
\label{fig:sketch_MS}
\end{figure*}

A sketch of a hit triplet in a uniform magnetic field is shown in \autoref{fig:sketch_MS}.
The $z$-axis of the spherical coordinate system is aligned with the magnetic field direction, and the $x$-$y$ plane is the bending plane.
In this plane, the line connecting hit~0 and~1 (hit~1  and~2) defines the azimuthal chord angle  $\varphi_{01}$ ($\varphi_{12}$), according to:
\begin{align}
  \varphi_{01} \, & := \;\sphericalangle_\varphi(\vec x_{1} - \vec x_{0}), \hspace{5mm}  \varphi_{12} \, := \;\sphericalangle_\varphi(\vec x_{2} - \vec x_{1}). \label{eq:chord_phi} 
\end{align}
The polar angles, $\theta_{01}$ and $\theta_{12}$, are defined in the longitudinal $s$-$z$ plane  as:
\begin{align}
  \theta_{01} \, & := \; \acot{ \left(\frac{z_{01}}{s_{01}} \right) }, \hspace{5mm}   \theta_{12} \, := \; \acot{ \left( \frac{z_{12}}{s_{12}} \right) },
\end{align}
with $s_{01}$ and  $s_{12}$ being the transverse arc lengths of the first and second triplet segment, respectively.
The corresponding bending angles, $\Phi_{01}$ and $\Phi_{12}$, are related to the transverse arc lengths via
\begin{align}
  \Phi_{01} \, & := \; \frac{s_{01}}{R_{01}} \, = \, s_{01} \; \wideparen{\curv_{\!\perp_{01}}} ,  \hspace{5mm}   \Phi_{12} \, := \; \frac{s_{12}}{R_{12}} \, = \, s_{12} \; \wideparen{\curv_{\!\perp_{12}}} ,
\end{align}
with $\wideparen{\curv_{\!\perp_{01}}}=\curv/\sin(\theta_{01})$ and $\wideparen{\curv_{\!\perp_{12}}}=\curv/\sin(\theta_{12})$ being the transverse curvatures of the two segments\footnote{
  Here and in the following, arcs indicate segment curvatures, which always have two indices indicating the connected hits.
}.

With above definitions, the kink angles in the bending and non-bending plane are given by\footnote{
Note that different symbols are used for denoting azimuthal angles: $\varphi$ is used to describe relative hit positions, $\phi$ is used to describe the track direction, and $\Phi$ is used for bending angles.
}:
\begin{align}
\Delta \phi  \;&= \; (\varphi_{12}-\varphi_{01}) - \frac{\Phi_{01} + \Phi_{12}}{2} \quad ,
  \label{eq:phikink} \\
\Delta \theta \;&=\; \theta_{12} - \theta_{01}
\quad .
  \label{eq:thetakink} 
\end{align}

In a uniform magnetic field, the bending angles $\Phi_{kk'}$ and polar angles $\theta_{kk'}$  fulfill the following relations~\cite{ref:MSPaper}:
\begin{align}
  \sin^2{\frac{\Phi_{kk'}}{2}} \;& = \; \frac{1}{4}\,  \curv^2 \,  d_{kk'}^2 \ + \ \curv^2 \, z_{kk'}^2 \, 
  \frac{\sin^2{(\Phi_{kk'}/2)}}{\Phi_{kk'}^2} \label{eq:condphi1} , \\
  \sin{\theta_{kk'}} \;& = \;\frac{1}{2}\; \curv \;  d_{kk'}  \; \cosec{\left( \frac{ \curv \, z_{kk'}}{2
    \cos{\theta_{kk'}}} \right)}  \label{eq:condtheta1} ,
\end{align}
where $k$ and $k'$ denote two consecutive hits that delimit a segment.

Above equations are transcendent and have multiple solutions, in general.
For the hit triplets considered here, only the solution with the smallest bending angle is relevant.
Instead of solving the equations above numerically,
\autoref{eq:condphi1} and \autoref{eq:condtheta1} are solved with the linearization ansatz introduced in \autoref{sec:triplet_parameters}.

For the linearization, the triplet trajectory with $\Delta \phi=0$ (no kink in the bending plane) is chosen as the reference and will be referred to as the \textsl{circle solution} and denoted by the superscript``C'' hereafter.
The circle solution serves as a suitable reference for linearization, provided that the MS angles are not too large.
The transverse curvature of the circle solution, $\curv_{\!\perp}^{\rm C}$, is readily obtained from the three triplet hit positions 0, 1, 2:
\begin{align}
  \curv_{\!\perp}^{\rm C} & = \; \frac{2 \; \sin(\varphi_{12}-\varphi_{01})}{d_{02}} \label{eq:circle1} \\
  & = \;
        2 \;  \frac{ [ ({\vec x}_1-{\vec x}_0) \times ({\vec x}_2-{\vec x}_1)]_z  }{d_{01}\; d_{12}\; d_{02}}
                      \label{eq:circle2} .
\end{align}
Here, 
$d_{kk'} := || ({\vec x}_{k_\perp} - {\vec x}_{k'_\perp}) ||$   are the hit distances in the transverse plane.
For the circle solution, the bending and polar angles are given by:
\begin{align}
\Phi_{kk'}^{\rm{C}} \; & := \; 2 \arcsin \left({\frac{d_{kk'}\;\curv_{\!\perp}^{\rm C}}{2}} \label{eq:phi_c1}\right)  , \\
  \cot \theta_{kk'}^{\rm C} \; &  := \; \frac{z_{kk'}}{d_{kk'}} \; \frac{\sin{(\Phi_{kk'}^{\rm{C}}/2)}}{\Phi_{kk'}^{\rm{C}}/2}
                               \, = \, \frac{z_{kk'} \, \curv_{\!\perp}^{\rm C}}{\Phi_{kk'}^{\rm{C}}}  \label{eq:theta_c1} ,
\end{align}
with indices $kk'=01$ for the first, and $kk'=12$ for the second segment.

By using the linearization ansatz of \autoref{eq:linearization_phi} and  \autoref{eq:linearization_theta},
 the four fundamental triplet parameters are eventually obtained:
\begin{align}
{\tilde \Phi} & = \;  \frac{1}{2} (\Phi_{01}^{\rm{C}} \; n_{01}^{\rm{C}}   +
  \Phi_{12}^{\rm{C}} \; n_{12}^{\rm{C}})   \label{eq:Phi0} , \\
  {\tilde \Theta} &  = \;  \theta_{12}^{\rm C} - \theta_{01}^{\rm C}
    \nonumber \\                & \quad
   \; + \;   (1-n_{12}^{\rm C}) \cot \theta_{12}^{\rm C} - (1-n_{01}^{\rm C}) \cot \theta_{01}^{\rm C}
                     \label{eq:Theta0}  ,  \\
  {\rho}_{\rm \phi} & = \;
- \frac{1}{2 \, \curv_{\!\perp}^{\rm C}} \left( \frac{\Phi_{01}^{\rm C} \; n_{01}^{\rm{C}}}{\sin \theta_{01}^{\rm C}} + \frac{\Phi_{12}^{\rm{C}}
                      \;   n_{12}^{\rm{C}}}{\sin \theta_{12}^{\rm C}} \right) , \label{eq:rho_phi} \\
{\rho}_{\rm \theta} & = \; \frac{1}{\curv_{\!\perp}^{\rm C}} \left(
                      (1-n_{01}^{\rm C}) \, { \frac{\cot \theta_{01}^{\rm C}}{\sin \theta_{01}^{\rm C}} } \;
                      - \; (1-n_{12}^{\rm C}) \, \frac{\cot \theta_{12}^{\rm C}}{\sin \theta_{12}^{\rm C}} \right)   .   \label{eq:rho_theta}
\end{align}
The fundamental triplet parameters depend on two index parameters, $n_{01}^{\rm{C}}$ and $n_{12}^{\rm{C}}$, which were first introduced in Ref.\cite{ref:MSPaper} and read\footnote{The index parameters were called $\alpha_1$ and  $\alpha_2$ in~\cite{ref:MSPaper}.
}:
\begin{align}
  n_{kk'}^{\rm C} \,&=\, \left({  \frac{\Phi_{kk'}^{\rm{C}}}{2}
                 \cot{\frac{\Phi_{kk'}^{\rm{C}}}{2}}\; \sin^2{\theta_{kk'}^{\rm C}} \; +\; \cos^2{\theta_{kk'}^{\rm C}} }\right)^{-1}
                  \label{eq:def_index_par} .
\end{align}
Note that the index parameters might become singular for recurling tracks, i.e., $|\Phi_{kk'}^{\rm{C}}| \ge \pi$.
In that case, the corresponding track segment provides an excellent momentum resolution as MS uncertainties vanish in first order~\cite{ref:MSPaper}.

\subsubsection{Small Bending Limit} \label{sec:smallbending}
In the limit of small bending angles, $\Phi_{kk'} \rightarrow  0$ (e.g. high momentum tracks)
 the index parameters approach unity, $n_{kk'}^{\rm C} \rightarrow 1 $.
The fundamental triplet parameters then simplify to:
\begin{align}
  \lim_{\curv \rightarrow 0} 
  {\tilde \Phi} & = \; \varphi_{12} - \varphi_{01}
 \label{eq:phi0_limit} ,
  \\
  \lim_{\curv \rightarrow 0} 
  {\tilde \Theta} & = \;  \theta_{2} - \theta_{1}
\label{eq:theta0_limit} ,
  \\
  \lim_{\curv \rightarrow 0} {\rho}_{\rm \phi} &= \;  - \frac{1}{2} \, \sqrt{d_{02}^2+z_{02}^2}
                                                   \; = \; - \frac{1}{2} || \vec x_{2} - \vec x_{0}   ||
\label{eq:rho_phi_limit} ,
  \\
  \lim_{\curv \rightarrow 0} {\rho}_{\rm \theta} & = \;  0 
\label{eq:rho_theta_limit} .
\end{align}
In this limit, the parameters ${\tilde \Theta}$ and ${\tilde \Phi}$ become the triplet kink angles,
the ${\rho}_{\rm \phi}$ parameter becomes half the negative chord length of the line connecting the first and third hit, and the ${\rho}_{\rm \theta}$ parameter vanishes.
All fundamental triplet parameters are very simple functions of the triplet geometry in this limit.

\subsection{Triplets in Zero Magnetic Field} \label{sec:zero_field}
In the case of zero magnetic field, the triplet trajectory is described by two straight lines with a kink.
Although the momentum cannot be measured via the Lorentz force (both triplet $\rho$-parameters are zero), the amount of MS at the middle tracking layer can be used as an indirect measure of the momentum.
The compatibility of the kink angles with MS theory can be tested using the relation:
\begin{align}
   \chi^2_{\textrm{MS}(B=0)} 
  & = \;   {\bs \Psi}^\top \bs D_\textrm{MS}(p) \, {\bs \Psi}  
        \label{eq:chi2_global_MS_no_B} .
\end{align}
Here hit position errors are neglected for the sake of simplicity.
Because of the momentum dependence of the MS errors: $  \sigma_{\rm MS} \propto {1}/{p}$, the right side of \autoref{eq:chi2_global_MS_no_B} can be rescaled:
\begin{align}
   \chi^2_{\textrm{MS}(B=0)} 
  & = \;   \frac{p^2}{p_0^2} \;  ({\bs \Psi}^\top \bs D_\textrm{MS}(p_0) \,  {\bs \Psi})  
        \label{eq:chi2_global_MS_no_B2} ,
\end{align}
where $p_0$ is a reference momentum that can be freely chosen.
Minimizing above equation is not a promising strategy as this would result in $p=0$, an unphysical solution.
A better approach is to ``calibrate'' the momentum such that the $\chi^2$-value per degree of freedom is~$1$.
This method works reliably for a large number of triplets (hits).

For each triplet, the momentum is the only degree of freedom.
After summing over all triplets, the expected mean value of the $\chi^2$
distribution is $\E[\chi^2] = n_\text{triplet}$ and the expected
variance is $\Var[\chi^2] = 2 n_\text{triplet}$.  
The best estimate for a momentum estimation is therefore given by the condition:
\begin{align}
   \chi^2_{\textrm{MS}(B=0)}   ~ \eqex  ~ n_\text{triplet} ,
\end{align}
resulting in:
\begin{align}
p_{\textrm{MS}(B=0)}    & = \;  p_0 \,  \sqrt \frac{n_\text{triplet}}{{\bs \Psi}^\top \,\bs D_\textrm{MS}(p_0) \;  {\bs \Psi}}
        \label{eq:momentum_global_MS_no_B2} , \\
  \sigma(p)_{\textrm{MS}(B=0)}^2 & = \;  p_0^2 \, \frac{2 \, n_\text{triplet} }{{\bs \Psi}^\top \,\bs D_\textrm{MS}(p_0) \;  {\bs \Psi}}
        \label{eq:sigma_global_MS_no_B2} .
\end{align}
Both equations can be combined to:
\begin{align}
  \sigma(p)_{\textrm{MS}(B=0)}^2 & = \;   \, \frac{2 \, p_{\textrm{MS}(B=0)}^2}{n_\text{triplet}} .
\end{align}
As  expected, the relative momentum resolution $\sigma(p)/p$ improves with $1/\sqrt{n_\text{triplet}}$.

In the presence of hit position errors, the same method can be used by making the following replacement (see also \autoref{eq:kmatrix_inverse}):
\begin{align}
 \bs D_\textrm{MS}(p_0)^{-1} & \rightarrow \; \bs K^{-1} \, =  \bs D_\textrm{MS}(p_0)^{-1} \; + \; \frac{p^2}{p_0^2} \,   \vec{\bs H} \Dvvb_\textrm{hit}^{-1}  \vec{\bs H}^\top
\end{align}
However, as the momentum determines the relative weight between MS and hit position errors, the solution needs to be determined iteratively.

Above method can be used for time projection chambers or spectrometers, where trackers are located outside the magnetic field region, like the LHCb experiment \cite{LHCb:2008vvz}.

\section{Energy Loss Correction} \label{sec:energy_loss}
Any tracking detector causes energy losses of particles through ionization.
For electrons and positrons, also Bremsstrahlung has to be considered.
The expected energy loss depends on the effective path length in the material, which is the same as used for the calculation of the MS parameters (\autoref{eq:def_b_ms}),  and is already known at triplet level.

For thin tracking layers, the corrections to the track fit are typically minor.
A simple method to consider small expected energy losses is discussed in \autoref{sec:eloss_local} for the local, and in \autoref{sec:energy_loss_correction} for the global track fit.
For thick tracking layers, however, energy straggling might be significant, requiring a combined fit of the track parameters and the energy losses, which is presented in \autoref{sec:energy_loss_fit}.

For the sake of simplicity, it is assumed that the particle is highly relativistic, such that the relation $p \sim E$ can be exploited.
Furthermore, the calculations below are performed for a uniform magnetic field.

\subsection{Energy Loss Correction in Local Fit} \label{sec:eloss_local}
\begin{figure}[tb]
   \begin{picture}(0.5\textwidth,110)
     \put(5,0){
       \epsfig{file=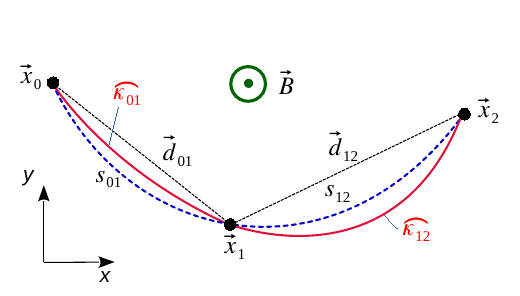,width=0.45\textwidth}
     }
   \end{picture}
   \caption{Sketch of a particle trajectory (\textsl{red solid} line) with energy loss at the middle triplet layer in a uniform magnetic field.
     The track curvatures of the segment before and after the energy loss are denoted by $\wideparen{\curvSa}$ and $\wideparen{\curvSb}$, respectively. For comparison a trajectory without energy loss is shown, which also connects all hits  (\textsl{blue dashed} line).  }
\label{fig:triplet_eloss}
\end{figure}
A hit triplet with hits $k=\{0, 1, 2\}$ is considered.
The energy loss at the middle tracking layer $\Delta_{E_1}$  changes the 3D curvature, $\curv$,
and the trajectory as sketched in \autoref{fig:triplet_eloss}.
Under the assumption of small energy losses,
${\Delta_{E_1}}  \ll q B/{\curvSa}  $, the curvature change is given by:

\begin{align}
  \Delta \curv_1 &= \; \wideparen{\curvSb} - \wideparen{\curvSa} \; = \; 
                           \frac{\Delta_{E_1}}{|{\vec p}|} \; \wideparen{\curvSa} \; \approx \; \frac{\Delta_{E_1}}{q B} \; \wideparen{\curvSa}^2 ,  \label{eq:curvature_shift_local}
\end{align}
with  $\wideparen{\curvSa}$ and $\wideparen{\curvSb}$ being the curvature of the first and second segment, respectively, and $\Delta \curv_1$ being the curvature change at the hit position $k=1$.
The curvatures before and after the energy loss are related to the solution of the triplet fit without energy loss, $\curv_0$, through:
\begin{align}
  \wideparen{\curvSa} \;&= \; \curv_0 \, - \,  \frac{s_{12}}{s_{02}}   \, \Delta \curv_1  \label{eq:c3d_energy_loss1} , \\
  \wideparen{\curvSb} \;&= \; \curv_0 \, + \,  \frac{s_{01}}{s_{02}} \, \Delta \curv_1    \label{eq:c3d_energy_loss2} .
\end{align}
Due to the curvature difference between the two segments, the azimuthal track angles are rotated with respect to the $\Delta \curv_1=0$ ($\Delta_{E_1}=0$) solution.
Using the notation from \autoref{sec:triplet_parameters}, the rotations in the bending plane at the three hit positions are given by:
\begin{align}
\Delta \phi_{01} &=\; \phi_{01}^{\Delta_{E_1}} - \phi_{01} \, =  \,   + \; \frac{s_{01} \, s_{12}}{2 \, s_{02}} \, \Delta \curv_1 \, \sin{\hat \theta} \label{eq:phi_rot_energy_loss1} , \\
\Delta \phi_{10} &=\; \phi_{10}^{\Delta_{E_1}} - \phi_{10} \, =  \, - \; \frac{s_{01} \, s_{12}}{2 \, s_{02}} \, \Delta \curv_1 \, \sin{\hat \theta} , \\
\Delta \phi_{12} &=\; \phi_{12}^{\Delta_{E_1}} - \phi_{12} \, =  \, - \; \frac{s_{01} \, s_{12}}{2 \, s_{02}} \, \Delta \curv_1 \, \sin{\hat \theta} , \\
\Delta \phi_{21} &=\; \phi_{21}^{\Delta_{E_1}} - \phi_{21} \, =  \, + \; \frac{s_{01} \, s_{12}}{2 \, s_{02}} \, \Delta \curv_1 \, \sin{\hat \theta} ,  \label{eq:phi_rot_energy_loss4}
\end{align}
with $\phi_{kk'}^{\Delta_{E_1}}$ and $\phi_{kk'}$ being the azimuthal track angles with and without energy loss at the middle hit position, respectively.
Note that  the angles  $\phi_{10}$ and $\phi_{12}$ change in the same way; thus the kink angles and the fit quality of the local fit are not affected by the energy loss, in first order.

\subsection{Energy Loss Correction in Global Fit} \label{sec:energy_loss_correction}
The expected energy losses at the detection layers are assumed to be known, for example by calculating the expected particle-specific ionization loss from the material distribution.
The energy losses are then treated as fixed parameters in the global fit.

The curvature of the  segment $\wideparen{  {\curv}_{kk'}}$
can be expressed by the curvature of the first segment, $\wideparen{{\curv}_{01}}$ and the sum of all curvature changes at the detector layers:
\begin{align}
\wideparen{  {\curv}_{kk'}} \, = \, \wideparen{{\curv}_{01}} \, + \, \sum_{\ell=1}^{k}  \Delta {\curv}_{\ell}  \label{eq:energy_conservation1} .
\end{align}
For triplet $k$, the relation between triplet curvature and segment curvature (compare \autoref{eq:c3d_energy_loss1}) then reads:
\begin{align}
  {\curv}_{j=k}  \; &= \;  \wideparen{\curv_{kk'}}  \, + \,  \frac{s_{k'k''}}{s_{kk''}}   \, \Delta \curv_{k'}  \label{eq:curvature_Eloss_correction},
\end{align}
where the convention $k''=k'+1$ and $k'=k+1$ is used.

Furthermore, an \emph{integrated energy loss} at the hit position $k$ is defined as:
\begin{align}
{I_{k}} &:= \; \sum_{\ell=1}^{k}  \Delta_{E_{\ell}}   \label{eq:def_integrated_energy} .
\end{align}
Using the approximation: 
\begin{align}
  \wideparen{\curv_{kk'}}   \; & \approx \; \wideparen{\curvSa} \; + \; \frac{I_{k}}{q B} \; {\wideparen{\curvSa}}^{\,2}, \label{eq:curvature_Eloss_quadratic}
\end{align}
the triplet curvature at triplet $j$ can be written as:
\begin{align}
{\curv}_{j=k}   \; & \approx \; \wideparen{\curvSa} \; + \; \frac{I^\star_{k}}{q B}  \; {\wideparen{\curvSa}}^{\,2} , \label{eq:triplet_Eloss_quadratic}
\end{align}
Here, $I^\star_{j}$ denotes an \emph{effective} integrated energy loss, defined by:
\begin{align}
  I_{k}^\star \; &= \;  I_{k} \, + \, \frac{s_{k'k''}}{s_{kk''}} \, \Delta_{E_{k'}} ,
\end{align}
which takes into account the energy loss inside the triplet $j$, see \autoref{eq:curvature_Eloss_correction}.

In this global fit, the first segment curvature $\wideparen{\curvSa}$ is chosen as fit parameter.
In order to avoid higher order terms of $\wideparen{\curvSa}$ in the fit, \autoref{eq:triplet_Eloss_quadratic} is linearized using the ansatz:
\begin{align}
  \wideparen{\curvSa}^2 \; &= \;  \hat \curv_j^2 \, + \, 2 (\wideparen{\curvSa} - \hat  \curv_j) \, + \, {\cal{O}}\left( (\wideparen{\curvSa} - \hat  \curv_j)^2 \right) ,
\end{align}
with $\hat \curv_j$  being the curvature obtained from the local triplet fit without energy loss.

\autoref{eq:triplet_Eloss_quadratic} in linearized from then reads:
\begin{align}
   {\curv}_{j}   \;  & \approx  \; \wideparen{\curvSa} \; \left(1+ \frac{2 \, I^\star_{j}}{q B} \, {\hat {\curv}_{j}}   \right) \, - \, \frac{I_{j}^\star}{q B} \, \hat {\curv}_{j}^{\,2} \ .     \label{eq:def_Istar} 
\end{align}
By comparing \autoref{eq:def_Istar} with the linearization ansatz in \autoref{eq:MS_theta_linearization} and \autoref{eq:MS_phi_linearization}, one sees that the energy loss inside a triplet corresponds to a change of the triplet parameters.
The linear relationship between ${\curv}_{j}$ and $\wideparen{\curvSa}$ enables the use of the global track fit formulas presented in \autoref{sec:global_track_fit} to include energy losses, by making the following substitutions:
\begin{align}
  \tilde \Theta_j \;  &  \rightarrow \ \tilde \Theta_j^\prime = \tilde \Theta_j  \, - \, \rho_{\theta,j}  \, \frac{I_{j}^\star}{q B} \, \hat {\curv}_{j}^2  \label{eq:energ_loss_substitution1} , \\
  \tilde \Phi_j \;  &  \rightarrow \ \tilde \Phi_j^\prime = \tilde \Phi_j  \, - \, \rho_{\phi,j}  \, \frac{I_{j}^\star}{q B} \, \hat {\curv}_{j}^2 ,  \\
  \rho_{\theta,j} \;  &  \rightarrow \ \rho_{\theta,j}^\prime = \rho_{\theta,j} \left(1 \, + \, \frac{2  \, I_{j}^\star}{q B} \, \hat {\curv}_{j} \right) , \\
  \rho_{\phi,j} \;  &  \rightarrow \ \rho_{\phi,j}^\prime = \rho_{\phi,j} \left(1 \, + \,  \frac{2  \, I_{j}^\star}{q B} \, \hat {\curv}_{j} \right)    \label{eq:energ_loss_substitution4} .
\end{align}
With above re-interpretation of the fundamental triplet parameters, energy corrections can easily be included in the global triplet track fit\footnote{
It is important to note that the locally fitted curvatures $\hat {\curv}_{j}$ should only be used 
in above equations if they are reliably measured by the triplets. 
An alternative strategy is to repeat the global fit, after neglecting the energy loss in the first step, and using the curvature result of the first fit as reference.
}.

\subsection{Combined Track and Energy Loss Fit} \label{sec:energy_loss_fit}
For thick tracking detectors, energy straggling might be significant, motivating to include energy losses in the tracking layers as additional fit parameters.
Energy losses have usually non-Gaussian tails.
However, for the sake of simplicity, a normal distribution is used to derive the fit formulas.

A difficulty arises from the quadratic curvature dependence of the curvature shifts (see e.g. \autoref{eq:curvature_shift_local}) that creates non-linearities in the fit\footnote{This problem does not arise if the 3D radius $R_\textrm{3D}=\curv$ is chosen as fit parameter, as done in Ref.\cite{ref:MSPaper}.}.
This problem can be tackled either by using the linearization ansatz from the previous section (\autoref{eq:energ_loss_substitution1} to~\ref{eq:energ_loss_substitution4}) or by re-iterating the fit.
In order not to add too much complexity to the discussion, it is assumed that the track curvature (momentum) is known well enough, either from the local triplets or a previous global fit, such that the curvature shifts can be approximated by:
\begin{align}
  {\curv}_{j}  - \wideparen{\curvSa} \; &\approx \;  \, \frac{I^\star_j}{q B}  \;
                                     {\curv}_\textrm{pre}^2 \quad , \label{eq:shift_curvature_eloss2}
\end{align}
with $\curv_\textrm{pre}$ being the curvature obtained pre-fit.

If hit position errors are neglected, the fit function  reads in matrix representation:
\begin{align}
&  \chi^2(\wideparen{\curvSa} , \, \bs  \varepsilon )   \  = \    { \bs \varepsilon}^\top  {{\bs D}}_\textrm{loss}  \;  { \bs  \varepsilon} \ +    \label{eq:chi2_energy_loss} \\
& \quad   \left( {\bs \Psi}  +  {\bs \rho} \, \wideparen{\curvSa}  - {\bs R} \,  \bs   \Delta_E \right)^\top \!
  \bs D_\textrm{MS} \,
     \left( {\bs \Psi}  +  {\bs \rho} \, \wideparen{\curvSa}  - {\bs R} \,  \bs  \Delta_E     \right)  ,
\nonumber
\end{align}
with $\wideparen{\curvSa}$ being the fit parameter describing the curvature of the first track segment,
$\bs \Delta_E$ being a vector describing the energy losses for each tracking layer, 
 $\bs  \varepsilon$  being a vector (fit parameter) describing the difference between the fitted energy losses and the expected energy losses, according to
$\bs \varepsilon := \bs \Delta_E - \bs \Delta_E^\textrm{exp} $, 
and ${{\bs D}}_\textrm{loss}$ being the energy loss precision matrix (inverse covariance matrix).

The relation between energy losses and   kink angle shifts is described by 
the matrix ${\bs R}$, which is  of size  $2 \, n_\text{triplet} \times n_\text{triplet} $, and given by:
\begin{align}
  \bs R \; &= \; \frac{{\curv}^2_\textrm{pre}}{q \, B} \left(
             \begin{array}{c}
               \diag{(\bs \rho_\theta)}   \; \bs \Sigma \\
               \diag{(\bs \rho_\phi)}   \; \bs \Sigma
             \end{array}
             \right) .
\end{align}
$\bs \Sigma$ is a quadratic \textit{integration matrix}, which sums up all energy losses before the respective tracking layer:
\begin{align}
  (\bs \Sigma)_{jj'} \, & = \,
                          \left\{
                          \begin{array}{c@{\quad\quad}c}
                          1   & \text{if} \quad j > j'  \\
     \frac{s_{j,j+1}}{s_{j-1,j+1}}   & \text{if} \quad j = j' \\
    0   & \text{if} \quad j < j'
                          \end{array}
                          \right. \ .
       \nonumber  
\end{align}
Note that the index $j$ runs over all triplets and that the energy loss in the very first tracking layer ($k=0$) and the last tracking layer ($k=n_\text{hit}-1$) are not accounted for in the fit.

The minimization of \autoref{eq:chi2_energy_loss} gives the result:
\begin{align}
  \curvloss  \;&=\;   - \frac{\bs \rho^\top  \bs K_\textrm{loss} \, \bs \Psi_\textrm{loss}} {\bs \rho^\top  \bs K_\textrm{loss}  \, \bs \rho}
                         \label{eq:ionization_fit_c3d} ,
  \\
  \sigma_{{\curv}_\textrm{loss}}^2 \;& = \;   \frac{1} {\bs \rho^\top  \bs K_\textrm{loss} \, \bs \rho} \label{eq:ionization_fit_sigma} ,
\end{align}
with $\bs K_\textrm{loss}$ being the triplet precision matrix for the energy loss fit, defined as:
\begin{align}
  \bs K_\textrm{loss}^{-1} \;& =\;     \bs D_\textrm{MS}^{-1} \; + \; {\bs R} \, \bs D_\textrm{loss}^{-1} \,  {\bs R}^\top ,  \label{eq:ionization_kmatrix_inverse}
\end{align}
and $\bs \Psi_\textrm{loss}$ being a modified kink angle vector, which includes  energy loss effects:
\begin{align}
  \bs \Psi_\textrm{loss} \;&= \; \bs \Psi \, + \, \bs R \, \bs \Delta_E^\textrm{exp} .
\end{align}
The fit quality then becomes:
\begin{align}
  \chi^2_\textrm{loss} \;& =\;     \bs \Psi_\textrm{loss}^\top  \bs \Krho_\textrm{loss} \bs \Psi_\textrm{loss}
  \label{eq:ionization_fit_chi2} ,
\end{align}
with
\begin{align}
\bs  \Krho_\textrm{loss}  \;&=\;  \left( \bs K_\textrm{loss}   - \frac{\bs K_\textrm{loss} \, \bs \rho \bs \rho^\top \bs K_\textrm{loss}}  { \bs \rho^\top \bs K_\textrm{loss} \,  \bs \rho}   \right)  .  \label{eq:ionization_Krho}
\end{align}
Above results are identical to the general fit that includes hit position errors when the hit residuals are replaced by the energy losses, and the matrix $ \vec {\bs H}$ is replaced by the matrix~$\bs R$.
Accordingly, one obtains for the fitted energy loss vector:
\begin{align}
  \bs \Delta_{E}  \;&=\;   \bs \Delta_{E}^\textrm{exp} \, + \,   \bs D_\textrm{loss}^{-1} \, {\bs R}^\top  \; \bs  \Krho_\textrm{loss} \; \bs \Psi_\textrm{loss} \label{eq:ionization_fit_delta} ,
\end{align}
 and for the corresponding covariance matrix:
\begin{align}
  \textrm{\bf Cov}_{\bs \varepsilon} \;& = \;  \bs D_\textrm{loss}^{-1}  - \; \bs D_\textrm{loss}^{-1} \,  {\bs R}^\top  \bs  \Krho_\textrm{loss} \; \bs R \; \bs D_\textrm{loss}^{-1}   .        \label{eq:ionization_fit_cov}
\end{align}
It is straightforward to also include hit errors in the fit.

\section{Triplet Fit Parallelization and Computational Effort}  \label{sec:parallel}
  The approach of splitting a set of hits into triplets primarily aims to enhance the parallelizability of track fitting and track reconstruction programs, which is especially important for high-rate experiments where the track reconstruction is rate-limited.
The \GTTF\ has three steps: the calculation of the triplet parameters, the local triplet fit (optional filtering step), and the global triplet track fit.
The possible savings in computational effort through parallelization for each step are discussed below.

\paragraph{Triplet Parameters and Hit Gradients}
Two main cases can be distinguished.
In the most general case (e.g.~inhomogeneous magnetic field) the triplet parameters need to be derived from at least $2 \times 4$ track extrapolations per triplet (see \autoref{sec:GeneralFieldConfiguration} for details), which
are \textit{independent} and can be  parallelized.

In case that an analytical solution for the triplet parameters exists (e.g.~uniform magnetic field, see \autoref{sec:HomogeneousMagneticField}), computationally expensive track extrapolations are not required.
The determination of the up to $3 \times 3$ hit gradients then only involves simple geometrical (re-)calculations.
If MS errors dominate,  the calculation of the hit gradients is not required.
Since the triplet parameters of all triplets are independent, the triplet parameter calculation can be fully parallelized.
Note that in track reconstruction tasks, each triplet parameter needs to be calculated only once, even if it belongs to several track candidates.
This step can significantly profit from parallel hardware architectures like GPUs.

\paragraph{Local Triplet Fit as Filtering Step}
Hit triplets are often used as seeds in track reconstruction; an early filtering step can be very useful to reduce hit combinatorics and speed up processing time.
The filtering step  consists of calculating the local triplet fit quality (\autoref{eq:local_chi2_general})  and applying a quality cut.
Only for accepted triplets, the momentum and its error need to be calculated to enable checking the consistency of the momentum with other triplets.
This step can also be fully parallelized.

\paragraph{Global Triplet Track Fit}
In the general case, where hit position errors cannot be neglected,
the inversion of the triplet covariance matrix, $\bs K^{-1}$, is the most time-consuming step, if the number of tracking layers is large.
The limited ability to accelerate this step highlights the importance of the local triplet fit as an early filtering step.
The situation is different in case of dominant MS errors, where $\bs K^{-1}$ is diagonal, and full parallelization is possible.
The global fit quality, the track curvature, and its error can then be easily calculated from simple sums, see Equations~\ref{eq:global_MS_c3d} to~\ref{eq:global_MS_chi2}.
The computational effort is then marginal and the required calculations can be efficiently implemented on parallel hardware architectures like GPUs. \\

Modern silicon pixel detectors offer superb spatial resolution; and high-momentum particles -- for which hit position uncertainties dominate -- are relatively rare in hadron collider experiments.
As a result, the GTTF can be implemented in an almost fully parallelized manner.
An interesting aspect is the collection of filtered triplets and their subsequent combination into track candidates, ultimately forming a graph whose size depends on the purity of the filtered triplets.
Resolving triplet (hit) ambiguities is actually equivalent to disconnecting the graphs~\cite{ref:Diestel},
a task that can be efficiently  tackled by GNNs~\cite{ref:GNN} and cellular automata~\cite{ref:CA}.
The latter has been studied in a recent work in combination with the GTTF~\cite{ref:Gupta}.

\section{Tracking Regimes} \label{sec:regimes}

The \GTTF\ provides closed-form expressions for the track parameters and the covariance matrix, enabling an easy identification of the dominant sources of the track parameter uncertainties, and an assessment of the measurement accuracy of the triplet.
By using a classification scheme, which is described below, different  \textit{tracking regimes}  can be defined.
For the classification, two new quantities are introduced: the \textit{tracking scale parameter} and the \textit{curvature significance parameter}.
Both parameters      provide important input for accelerating track reconstruction, see discussion in \autoref{sec:parallel},  and can also aid in optimizing tracking detector designs.
}

\subsection{The Tracking  Scale Parameter}
The track parameter uncertainties are determined by the triplet precision matrix $\bs K$ (\autoref{eq:kmatrix_inverse}), which depends on the triplet uncertainties  $\GamPhi^\star$ and $\GamTheta^\star$ (\autoref{eq:gamtheta} and \autoref{eq:gamphi}).
These parameters combine hit position and MS errors.

For each triplet, the relative fraction of the hit position errors (variance) can be used to define two tracking scale parameters:
\begin{align}
\mu_\phi^2 \; := \; \frac{\GamPhi}{\GamPhi^\star}, \hspace{1cm}
\mu_\theta^2 \; := \; \frac{\GamTheta}{\GamTheta^\star}. \label{eq:def_tracking_scale}
\end{align}
The values of $\mu_\phi$ and $\mu_\theta$ are in the range from $0$ for dominant MS errors (low momentum regime) to $1$ for dominant hit position errors (high momentum regime).

The tracking scale parameters are of practical relevance for track fitting.
In the case that the tracking scale parameters are small ($\mu_{\phi,\theta} \lesssim 0.15$), the fast MS fit can be used\footnote{For $\rho_\phi^2 \gg \rho_\theta^2$ (small bending), only the tracking scale parameter $\mu_\phi$ is relevant.},
whereas in all other cases the general fit should be used, which requires the inversion of the triplet covariance matrix $\bs K^{-1}$.
The tracking scale parameter thus defines a criterion for determining whether the track fit can be accelerated.

\subsection{The Curvature Significance  Parameter}
The curvature significance parameter, $\xi$, is defined by the ratio of the curvature over its error:
\begin{align}
\xi^2 \; := \;  \frac{\curv^2}{\sigma_{\curv}^2}    \label{eq:def_curvature_significance} ,
\end{align}
and quantifies how precisely the track curvature is measured.
The parameter allows to distinguish between the strong bending regime, $\xi^2 \gg 1$, where a precise momentum measurement is possible, and the weak bending regime, $\xi^2 \approx 0$, where no momentum measurement is possible.

\subsubsection{Dominant MS Uncertainties ($\mu^2=0$)}
For a tracking detector in a uniform magnetic field,
the following relation is obtained if MS uncertainties dominate:
\begin{align}
  \xi_\textrm{MS}^2 \; & = \; 
\frac{\curvMS^2}{\sigma_{\curvMS}^2} \; \approx \; \frac{\tilde \Phi^2}{\sigma_{\phi_\textrm{MS}}^2}  .  \label{eq:curvature_significance_MS1}
\end{align}
In the small bending limit (\autoref{sec:smallbending}), the last expression can be rewritten  as:
\begin{align}
  \xi_\textrm{MS}^2 \; &   \; \approx \; \frac{a^2}{4 \, b_\textrm{MS}^2}  \label{eq:curvature_significance_MS2},
\end{align}
with $a$ being the length of the triplet in Euclidean space (compare \autoref{eq:rho_phi_limit}) and $b_\textrm{MS}$ being the  MS parameter (\autoref{eq:sigmams_c}).
$\xi_\textrm{MS}$ is independent of the particle momentum
and characterizes the tracking detector quality for MS dominance.
The requirement to measure the track curvature with $3\sigma$ significance corresponds to the condition
$\frac{a}{2 \, b_\textrm{MS}} \gtrsim 3$, a criterion first formulated in Ref.\cite{ref:MSPaper}.

It should be remarked that $\xi_\textrm{MS}$ is also related to the fitting bias discussed in \autoref{sec:Biases} and that for $\xi_\textrm{MS}<10$ significant  biases ($>\SI{1}{\percent}$) of the momentum (curvature) can occur.

\subsubsection{General case}
For non-negligible hit position errors (general case), the curvature significance parameter (\autoref{eq:def_curvature_significance}) can be generalized by including the tracking scale parameter (\autoref{eq:def_tracking_scale}):
\begin{align}
  \xi \; & = \; {\xi_\textrm{MS}} \cdot {\sqrt{1-\mu_\phi^2}}.
\end{align}
As expected, the curvature significance decreases as the contribution of hit position errors increases (larger $\mu_\phi$), which corresponds to higher track momenta.

\subsection{Examples of Tracking Regimes in Pixel Detectors}
\begin{figure}[tb]
   \begin{picture}(150,250)
     \put(-10,0){
       \epsfig{file=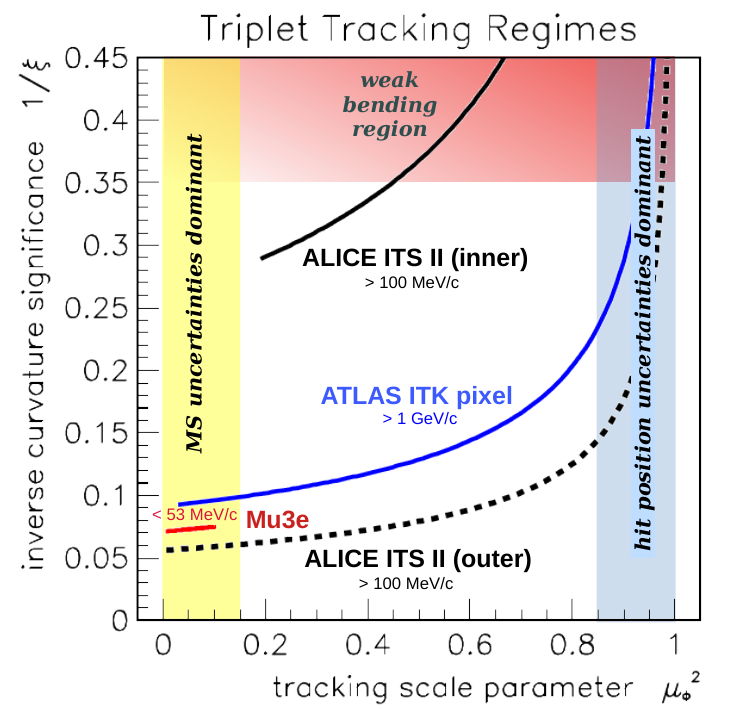,width=0.5\textwidth}
     }
   \end{picture}
   \caption{Parameter space spanned by the normalized MS parameter scale $\xi$ and the tracking scale parameter $\mu_\phi$ for various pixel tracking detectors: ALICE ITS II inner and outer pixel tracking detector (\textsl{black}), ATLAS ITK (\textsl{blue})
     and Mu3e (\textsl{red}).
     The curves cover the momentum range \SI{0.1} - \SI{100}{GeV/c} for ALICE, \SI{1} - \SI{1000}{GeV/c} for ATLAS, and \si{10} - \SI{53}{MeV/c} for Mu3e.
   }
\label{fig:fits_phase_space}
\end{figure}

For illustration,  the  $\mu_\phi$ and  $\xi$ parameters are calculated for several pixel detectors: the upgraded ALICE experiment~\cite{ref:ALICE_tracker,ALICE:2013nwm}, the upgraded ATLAS experiment~\cite{ref:ITK_Pixel_TDR}, and the Mu3e experiment~\cite{Mu3e:2020gyw}.
These detectors are here selected as they cover a wide range of particle momenta from {$\cal O$}(\SI{10}{MeV/c}) at the Mu3e experiment up to {$\cal O$}(\SI{1000}{GeV/c}) at the ATLAS experiment.
Based on the detector description in the respective technical design reports, the tracking regime parameters are derived from hit triplets in the central tracking regions.
The $\mu_\phi$ - $1/\xi$ parameter space for the above mentioned pixel trackers is shown in \autoref{fig:fits_phase_space}, and discussed below.

\paragraph{ALICE Inner Tracking System (ITS~II)}
The ITS~II detector~\cite{ref:ALICE_tracker,ALICE:2013nwm} is in operation since 2024.
It consists of three inner pixel layers with a radiation length of $X/X_0=\SI{0.35}{\percent}$ and four outer pixel layers with a radiation length of $X/X_0=\SI{1}{\percent}$.
The distance between the inner pixel layers is only $\sim \SI{10}{mm}$, which, together with the relatively moderate field of $B=\SI{0.5}{\tesla}$, results in  a low curvature significance of $\xi_\mathrm{MS} \sim 3.8$, despite the small amount of tracking material.
With the Inner ITS~II detector alone,
a $3\sigma$ measurement of track curvatures is  only possible for very low momenta ($p \lesssim \SI{150}{MeV/c}$).
However, it should be mentioned that the main purpose of the Inner ITS~II detector is vertexing and not the momentum measurement.

The situation is  different for the Outer ITS~II detector, where the tracking layers are separated by about $\SI{80}{mm}$.
The curvature significance for low-momentum tracks is $\xi_\mathrm{MS} \sim 18$ and a $3\sigma$ measurement of track curvatures is possible for a single triplet up to $\SI{15}{GeV/c}$, thanks to the high resolution of the ALPIDE sensors~\cite{ref:ALPIDE}.

\paragraph{ATLAS Inner TracKer (ITK)}
The ITK tracking system\cite{ref:ITK_Pixel_TDR} of the high luminosity upgraded ATLAS experiment was optimized to reconstruct charged particles at very high particle rates, with up to 200 collisions per bunch crossing.
Similar to the Inner ITS~II detector, the main purpose of the ATLAS pixel detector is to reconstruct primary and secondary vertices.
Due to the high radiation length  of about $X/X_0=\SI{1.5}{\percent}$ per tracking layer in the central region, a curvature significance of $\xi_\mathrm{MS} \sim 11$ is only reached for very low-momentum tracks.
A $3\sigma$ measurement of the curvature is possible for a single triplet up to $\SI{20}{GeV/c}$, thanks to the high field of $B=\SI{2}{\tesla}$.

\paragraph{Mu3e Pixel Tracker}
The Mu3e pixel detector~\cite{Mu3e:2020gyw} has an ultra-light pixel detector design with a radiation length of about $X/X_0=\SI{0.11}{\percent}$ per tracking layer, and was optimized for tracking low-momentum electrons and positrons from muon decays in the momentum range $12 - \SI{53}{MeV/c}$.
The tracking scale parameter is $\mu^2_\phi \le 0.1$ for all momenta, such that the fast MS track fit can be exploited for all tracks.
The curvature significance is about $\xi \sim 14$ in the full momentum range.
Note that the Mu3e pixel detector exploits recurling tracks, which provide an about 7 times better momentum resolution, corresponding to $\xi \sim 100$. \\

  The above examples show how the tracking scale parameter $\mu_\phi$ and the curvature significance $\xi$ distinguish different tracking regimes, which would require different implementations to accelerate the GTTF.
  Furthermore, the curvature significance also defines a metric for the ability to resolve inevitable hit ambiguities in the track reconstruction: triplets with large  $\xi$-values restrict the phase space more than low $\xi$ triplets.
For track reconstruction,  high $\xi$ triplets are better suited than low $\xi$ triplets as they provide higher precision in track extrapolations and more stringent consistency checks (e.g. consistency of triplet momenta).
Consequently, the proposed tracking regime analysis can serve as a valuable tool for track reconstruction optimization.

\section{Summary} \label{sec:summary}

In this paper a new track fit, the General Triplet Track Fit (\GTTF) was presented.
The solution of the fit is given in an analytical closed-form.
All formulas are based on so-called triplet parameters which contain the detector-specific information such as geometry,  detector material and the magnetic field.
The fit is generic and takes into account material effects such as MS and energy loss, as well as hit position  uncertainties.
The triplet representation  makes the  track fit universal, since the same fitting code can be used for all kind of tracking detectors.

Triplet parameters  are derived for various tracking detector configurations.
It is shown that in the case of small bending (high momentum tracks), the triplet parameters become simple geometrical constants.
For the triplet parameter calculation in a general (inhomogeneous) magnetic field, an algorithm employing track extrapolation is presented.
Furthermore, two methods for including energy losses are described: one that treats the expected energy loss as an additional input (suitable for small energy losses), and one that fits the energy loss for each tracking layer (suitable for large energy losses).

  The output of the \GTTF\  consists of the particle momentum (3D curvature), all hit residuals, and the full covariance matrix.
  From this, the state vector of the track, which is needed for track extrapolation (vertexing), can be calculated at any point of the trajectory. 
  Formulas to calculate the state vector and the corresponding covariance matrix  are provided in the appendix.

The \GTTF\ can be fully parallelized on triplet level.
Due to the high degree of parallelization, the \GTTF\ is ideal for its implementation on parallel hardware architectures such as GPUs.
Furthermore, the fit of a single hit triplet is over-constrained, thus making it possible to calculate a fit quality on triplet level and to apply filters.
This (optional) filtering step offers the possibility to accelerate track reconstruction at an early stage.

The GTTF provides several options for algorithmic acceleration, the most important being the MS fit,
which does not require a computationally expensive matrix inversion, in contrast to the global fit.
Two new parameters are introduced, the \emph{tracking scale parameter} and the \emph{curvature significance parameter}, to classify different tracking regimes, for which different track fit approximations and optimizations apply.

An interesting application for the tracking regime-spe\-ci\-fic track fit acceleration is track reconstruction at hadron colliders.
Since most particles produced in hadron interactions are at low momenta and dominated by MS,
significant amount of computating time can be saved by performing the fast MS fit.
It is recommended to execute the more time-expensive generic fit only if hit position errors cannot be neglected ($\mu_{\phi,\theta}\gtrsim 0.15$). 
An optimization of the track reconstruction based on the tracking regime concept is of high relevance for real-time applications, in particular in environments with high particle rates.
The analysis of tracking regimes can also provide useful information for the design of new tracking detectors.

The analytical closed-form of the GTTF also enables the study of the inherent fitting bias in the MS tracking model.
A detailed study of the bias in MS dominated track fit is performed.
Mitigation strategies are discussed and a regularized track fit is presented, which reduces  track fit biases.

Finally, it is important to note that the GTTF can be easily extended to incorporate geometric alignment parameters, which will be addressed in a follow-up article.

\section*{Acknowledgments}
The author thanks several group members and colleagues for many useful discussions on this topic and for proofreading: Niklaus Berger, Sebastian Dittmeier, David Fritz, Tamasi Kar, Alexandr Kozlinskiy, Abhirikshma Nan\-di, and Christof Sauer.


\appendix

\section{Global Fit for Dominant Hit Position Errors} \label{app:hit_position_errors}

In the limit of dominant hit position errors, the MS errors can be neglected.
According to \autoref{eq:kmatrix_inverse}, the precision matrix becomes:
\begin{align}
\bs K \rightarrow \bs K_\textrm{hit} \; := \; (\vec {\bs H} \Dvvb_\textrm{hit}^{-1} \vec {\bs H}^\top)^{-1} .
\end{align}
The 3D curvature and its uncertainties are then given by: 
\begin{align}
  \curvhit     
  \;&=\;  - \frac{{\bs \rho}^\top \bs K_\textrm{hit} \,  {\bs \Psi}}
                {{\bs \rho}^\top \bs K_\textrm{hit} \,  {\bs \rho}}  \label{eq:glob_hit_c3d} , \\
  \sigma_{\curv_\textrm{hit}}^2
  \;&=\;   \frac{1}
      {{\bs \rho}^\top \bs K_\textrm{hit} \,  {\bs \rho}}  \label{eq:glob_hit_sigma_c3d} .
\end{align}
Furthermore, the hit positions shifts and the corresponding covariance matrix are given by:
\begin{align}
  \bs {\vec \delta}_\textrm{hit}    
    \;&=\;  \Dvvb_\textrm{hit}^{-1}  \vec{\bs H}^\top  \; \bs {\Krho}_\textrm{hit} \; \bs \Psi  \label{eq:glob_hit_delta}  , \\
  \Covvvb_{\delta_\textrm{hit}} 
                              \;& = \;  \Dvvb_\textrm{hit}^{-1}  - \; \Dvvb_\textrm{hit}^{-1}  \vec{\bs H}^\top   \bs \Krho_\textrm{hit} \; \vec{\bs H} \; \Dvvb_\textrm{hit}^{-1}  \label{eq:glob_hit_cov} .
\end{align}
Here, $\bs  \Krho_\textrm{hit}$ is defined similar to \autoref{eq:global_fit_cov}:
\begin{align}
\bs \Krho_\textrm{hit}  \;&=\;  \left( \bs K_\textrm{hit}   - \frac{\bs K_\textrm{hit} \; \bs \rho \bs \rho^\top \bs K_\textrm{hit}}   { \bs \rho^\top \bs K_\textrm{hit} \; \bs \rho}   \right)  .  \nonumber
\end{align}
Finally, the fit quality is given by:
\begin{align}
  \chi^2_\textrm{hit} \;& = \; {\bs \delta}^\top \Dvvb_\textrm{hit} \, {\bs \delta}  \label{eq:glob_hit_chi2_1} \\
 \;& = \;    \bs \Psi^\top  \bs \Krho_\textrm{hit} \bs \Psi  .  \label{eq:glob_hit_chi2_2}
\end{align}
Note that in the first line of \autoref{eq:glob_hit_chi2_1} the sums are executed over all hit uncertainty directions whereas in the second line (\autoref{eq:glob_hit_chi2_2}) the sums run over twice the number of triplets.


\section{Regularized Local MS Fit} \label{sec:regularized_local}
The formulas for the regularized global MS fit are discussed in \autoref{sec:regularized_fit}.
For a single triplet, the fit result is given by:
\begin{align}
  {\curvMSreg}  \,& =\, - \; \frac{{\tilde \Phi}^2 \, \sin^2{\hat \theta}  + {\tilde \Theta}^2}
                    {{\tilde \Phi} \; \rho_\phi \,  \sin^2{\hat \theta}+{\tilde \Theta} \; \rho_\theta} ,
                       \\
  \sigma_{\curvMSreg} \,&=\,
  {b_\textrm{MS}}\;
  \frac{({\tilde \Phi}^2 \,
  \sin^2{\hat \theta}  + {\tilde \Theta}^2)^\frac{3}{2}}{({\tilde \Phi} \; \rho_\phi \,
  \sin^2{\hat \theta}+{\tilde \Theta} \; \rho_\theta)^2}  , \\
  \chi_\textrm{MSreg}^2                                         \,& =\,
  \frac{1}{b_\textrm{MS}^{2}}\;
     \frac{({\tilde \Phi} \; \rho_\theta - {\tilde \Theta} \; \rho_\phi)^2}
     {{\tilde \Phi}^2   + {\tilde \Theta}^2 /  \sin^2{\hat \theta}} .
\end{align}

\section{Strip Detectors} \label{app:strip_detectors}
In this section, the local triplet fit is discussed for a barrel-type strip detector in a uniform magnetic field.
In the context of the triplet fit, a detector is defined to be a strip detector if large hit position errors in one detector direction create a significant rotational triplet uncertainty.
In the following formulas for the local triplet fit are derived, which include the hit position dependence of the triplet parameters that has been neglected in \autoref{sec:global_track_fit} and \autoref{sec:local_fits}.
Rotational uncertainties in the global triplet fit  are usually negligible if several triplets are combined.

The configuration with strips oriented in axial direction is considered here, which is used in many experiments.
Since $\sigma_z \gg \sigma_\phi$, such a detector is not able to precisely measure the polar angle.
This, however, affects the 3D curvature via the relation $\curv = \curv_\perp \, \sin \theta$.

With the assumptions $\GamTheta \gg \sigma_\textrm{MS}^2$ and $\GamTheta \gg  \GamPhi$, \autoref{eq:local_c3d_general} (and the following equations)
can be approximated as:
\begin{eqnarray}
  {\curv}_{z\text{-strip}} & \approx  & - \frac{ {\tilde \Phi}  }
                             { \rho_\phi} \; + \;  \Delta {\curv}_\textrm{rot} , \label{eq:c3d_general_con_high_barrel} \\
  \sigma_{{\curv,z\text{-strip}}}^2 & \approx &
            \frac{ \GamPhi^\star } {\rho_\phi^2 }  \; + \;   \Delta \sigma^2_{\curv,\textrm{rot}} ,
                                                        \label{eq:sig_c_general_con_high_barrel} \\ 
  \chi^2_{z\text{-strip}}  &  \approx &  \frac{{\tilde \Theta}^2 }{\GamTheta} ,
  \label{eq:chi2_general_high_barrel}
\end{eqnarray}
where additional correction term are added to account for the rotational uncertainty.
They are calculated as:
\begin{eqnarray}
  \Delta {\curv}_\textrm{rot} &=& - \frac{ {\tilde \Phi}  }   { \rho_\phi} \; \Delta \theta_\text{rot}  \cot{\hat \theta} , \\
  \Delta \sigma^2_{{\curv}_\textrm{rot}} &=&  {\curv}_{z\text{-strip}}^2 \;   \sigma_{\theta_\text{rot}}^2 \,  \cot{\hat \theta} ,
\end{eqnarray}
and depend on the triplet rotation and its error:
\begin{eqnarray}
  \Delta \theta_\text{rot} & = & \frac{\tilde \Theta}{2 \GamTheta} \left(  {\vec \delta}_2 \, {\vec h}_{\theta_2} -   {\vec \delta}_0 \, {\vec h}_{\theta_0}  \right), \label{eq:polar_angle_rotation} \\
  \sigma_{\theta_\text{rot}}^2 & = &   \frac{1}{4} \, (  {{\vec h}_{\theta_0}}^t \Vvv^\prime \,  {{\vec h}_{\theta_0}} + {{\vec h}_{\theta_2}}^t \Vvv^\prime \,  {{\vec h}_{\theta_2}}   ) \ .
\end{eqnarray}


\section{Track Parameters and Covariance Matrix} \label{app:track_extrapol}
In a broken line fit such as the GTTF, kinks in the trajectory must be taken into in the determination of the track parameters.
In the following it is assumed that all detector material in the active tracking region is located at the tracking layers.
The position of the kinks thus agrees with the position of the hits.
For vertexing (extrapolation to the track origin), the track parameters need to be determined at the first hit ($k=0$).
For track extrapolation in the other direction, the track parameters need to be determined at the last hit ($k=n_\textrm{hit}-1$).
Note that for track extrapolation the material in the first (last) detector layer must be considered as extra scatterer (like any other material, e.g. beam pipes) since this material is not included in the track fit.

For track parameterization, the following representation is used:
\begin{align}
  {\bar t} = (\vec x, \curv, \theta, \phi) , \label{eq:track_para_def}
\end{align}
with the corresponding covariance matrix defined as:
\begin{align}
  \Covhh_{t} =
\left( \begin{array}{cccc}
  \Covvv (\vec x,\vec x)        & \Cov(\curv,\vec x) & \Cov(\theta,\vec x) & \Cov(\phi,\vec x) \\
  \Cov(\vec x,\curv)  & \Var(\curv)        & \Cov(\theta,\curv)  & \Cov(\phi,\curv) \\
  \Cov(\vec x,\theta) & \Cov(\curv,\theta) & \Var(\theta)        & \Cov(\phi,\theta) \\
  \Cov(\vec x,\phi)   & \Cov(\curv,\phi)   & \Cov(\theta,\phi)   & \Var(\phi) 
\end{array} \right) 
\end{align}
Here and below, the bar $\bar ~$ indicates a vector and the double bar $\bar \bar ~$ a matrix in the six-dimensional track parameter space.

\subsection{Track  Position and Uncertainties} \label{sec:track_position}
Be hit~$k$ the starting point of the track extrapolation, the track position in global coordinates
is obtained from the fitted residuals (\autoref{eq:global_fit_delta}) via the relation:
\begin{align}
  \vec x_{k,\textrm{fit}} = \vec x_{k}
  + {\vec {\vec Q}}_{k}^{\top} \, \vec \delta_{k} ,  \label{eq:fitted_hit_position}
\end{align}
with ${\vec {\vec Q}}_k^{\top}$ being the back-rotation from local to global coordinates, see
also \autoref{eq:transformation}.
The corresponding covariance matrix in global coordinates is derived from the local covariance matrix via the transformation: 
\begin{align}
  \Covvv (\vec x_k,\vec x_k) \, = \, (\Covvvb_{x})_k & =  \; (\Qvvb^\top  \, \Covvvb_{\delta} \;  \Qvvb)_k  , \label{eq:error_matrix_trafo}
\end{align}
where $\Covvvb_{\delta}$ is given by:
\begin{align}
 \Covvvb_{\delta} \;& = \;  \Dvvb_\textrm{hit}^{-1}  - \; \Dvvb_\textrm{hit}^{-1} \, \vec{\bs H}^\top   \bs  \Krho \; \vec{\bs H} \; \Dvvb_\textrm{hit}^{-1} 
   \tag{\ref{eq:global_fit_cov}}.
\end{align}

\subsection{Track Parameter Representation}
The track vector in local coordinate transformation, $\bar t_\textrm{loc}$, is defined equivalent to \autoref{eq:track_para_def}:
\begin{align}
\bar t_\textrm{loc} = (\vec \delta, \curv, \theta, \phi) .
\end{align}
For the transformation to global coordinates a $6 \times 6$ matrix is defined:
\begin{align}
\bar  {\bar G} & := \;
           \left(
           \begin{array}{cccc}
             \Qvv^\top  & 0 & 0 & 0 \\
             0 & 1 & 0 & 0 \\
             0 & 0 & 1 & 0 \\
             0 & 0 & 0 & 1 
           \end{array}
           \right)           .
\end{align}
The track parameters and the corresponding error matrix then transform as:
\begin{align}
  \bar t & =  \;   {\bar {\bar G}}^{-1} \, \bar t_\textrm{loc} \\
  \Covhh_{t} & =  \; \bar {\bar G} \; \Covhh_{t_\textrm{loc}} \;  {\bar {\bar G}}^{-1}  . \label{eq:error_track_trafo}
\end{align}
Note that the curvature and track angles are defined in global coordinates and not transformed.

\subsection{Track Curvature}
For the general case, the 3D curvature and its variance are given by (see \autoref{sec:global_track_fit}):
\begin{align}
  \curv \;&=\;   - \frac{\bs \rho^\top  \bs K \bs \Psi} {\bs \rho^\top  \bs K \bs \rho} , \tag{\ref{eq:global_fit_c3d}} \\
  \Var(\curv) \;& = \;   \frac{1} {\bs \rho^\top  \bs K \bs \rho} \tag{\ref{eq:global_fit_sigma}} .
\end{align}
The correlated error between the fitted curvature and the hit position $k$  is given by:
\begin{align}
  \Cov (\vec \delta_k,\curv) \;& = \; (\Covvb_{\delta/\curv})_k ,
\end{align}
with
\begin{align}
  \Covvb_{\delta/\curv} \;& = \; \frac{ \Dvvb_\textrm{hit}^{-1}   \vec{\bs H}^\top   {\bs K} \; {\bs \rho}}
                                      {{\bs \rho}^\top {\bs K} \; {\bs \rho}}  \label{eq:global_fit_cov_mixed}.
\end{align}

\subsection{Track Direction}
The determination of the track direction is more evolved since the track direction is position dependent in a magnetic field.
For an inhomogeneous magnetic field, the track direction can be determined using the method described in \autoref{sec:GeneralFieldConfiguration}.

\begin{figure}[tb]
   \begin{picture}(1.0\textwidth,160)
     \put(0,0){
       \epsfig{file=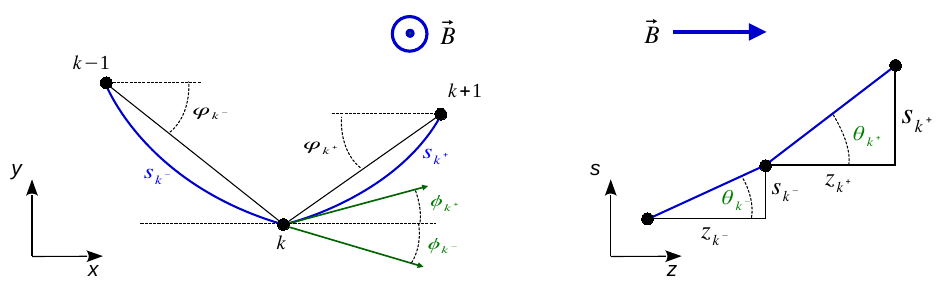,width=0.99\textwidth}
     }
   \end{picture}
   \caption{ Azimuthal (\textsl{left}) and polar (\textsl{right}) track angles at the hit position $k$ using the short-hand notation introduced in \autoref{eq:azimuth_hit_position} and \autoref{eq:theta_hit_position}.
     In a uniform magnetic field,
     the arc lengths are related to the bending angles via $s_{k^\pm}=\Phi_{k^\pm} R_{k^\pm}$, with $R_{k^\pm}$ being the bending radius.
}
\label{fig:drawing_triplet_direction}
\end{figure}

For a uniform magnetic field (\autoref{sec:HomogeneousMagneticField})
the polar and azimuthal angles  of a trajectory  at hit position $k$ are given as follows:

\begin{align}
\theta_{k^\pm} \, & = \; \acot \left[ \frac{z_{k^\pm}}{\;d_{k^\pm}} \; \frac{\sin{(\Phi_{k^\pm}/2)}} 
                    {\Phi_{k^\pm}/2}  \right]   \label{eq:theta_hit_position} , \\
  \phi_{k^\pm} \, & = \; \varphi_{k^\pm}  \mp \frac{\Phi_{k^\pm}}{2} \label{eq:azimuth_hit_position} ,
\end{align}
where a short-hand notation for the indices is used: $k^\pm = ``k,k \pm 1"$. The $\pm$ sign refers here to the two solutions after/before the scattering at tracking layer~$k$, see \autoref{fig:drawing_triplet_direction}.
Note that for the first (last) hit of a track only a solution with the $+$ ($-$) sign  exists and
that $\varphi_{k^\pm}$ is defined by  \autoref{eq:chord_phi}.
Further note that all parameters in \autoref{eq:theta_hit_position} and \autoref{eq:azimuth_hit_position} are post-fit.

\paragraph{Curvature Dependence}
Both \autoref{eq:theta_hit_position} and \autoref{eq:azimuth_hit_position} require knowledge of the bending angle $\Phi_{k^\pm}$, which in turn depends on the curvature $\curv$.
The GTTF consistently employs the small multiple scattering approximation, which assumes  that the triplet trajectory can be linearized around the circle solution in the bending plane (\autoref{eq:noMS}).

At segment level, the linearization of the polar and bending angle yields:
\begin{align}
  \theta(\curv)_{k^\pm}   \,
                   &         = \,  \theta_{k^\pm}^{\rm C} + \tau_{k^\pm} +   \curv \; \wideparen{{\rho_\theta}}_{k^\pm} \label{eq:seg_linearisation_theta}  , \\
  \Phi(\curv)_{k^\pm}   \, 
     & = \,  \nu_{k^\pm} \, + \, \curv \; \wideparen{{\rho_\phi}}_{k^\pm}     \label{eq:seg_linearisation_phi} , 
\end{align}
where new segment-specific linearization parameters are introduced:
\begin{align}
  \wideparen{\rho_{\theta}}_{k^\pm} & := \;  \frac{n_{k^\pm}^{\rm C}-1} {\wideparen{\curv}_{k^\pm}^{\rm C}} \, {\cot \theta_{k^\pm}^{\rm C}} , \label{eq:seg_lin_rho_theta} \\
  \wideparen{\rho_{\phi}}_{k^\pm} & := \;  \frac{n_{k^\pm}^{\rm C}} {\wideparen{\curv}_{k^\pm}^{\rm C}} \,    {\Phi_{k^\pm}^{\rm C}}  , \label{eq:seg_lin_rho_phi} \\
  \tau_{k^\pm} & := \; (1-n_{k^\pm}^{\rm C})  \, {\cot \theta_{k^\pm}^{\rm C}} , \label{eq:seg_lin_tau} \\
\nu_{k^\pm} & := \; (1-n_{k^\pm}^{\rm C})  \,  {\Phi_{k^\pm}^{\rm C}} , \label{eq:seg_lin_nu}
\end{align}
with $\wideparen{\curv}_{k^\pm}^{\rm C}$ being segment-wise 3D curvatures, given by:  
\begin{align}
\wideparen{\curv}_{k^\pm}^{\rm C} &    = \; \curv_{\perp_j}^{\rm C} \; \sin \theta_{k^\pm}^{\rm C} \label{eq:seg_lin_kappa}
\end{align}
As in \autoref{sec:HomogeneousMagneticField}, the index $C$ always refers to the circle solution.

The triplet index $j \in k-1,k$ indicates which of two possible triplets per segment is chosen  to determine the circle solution.
Note that the $\wideparen{~}$ is used here and below to distinguish the segment-wise defined $\wideparen{\rho}$ parameters (\autoref{eq:seg_lin_rho_theta} and \autoref{eq:seg_lin_rho_phi})  from the triplet-wise defined $\rho$ parameters (\autoref{eq:linearization_phi} and \autoref{eq:linearization_theta}).
Further note that the segment-wise $\wideparen{\rho}$ parameters are related to the triplet $\rho$ parameters (see also \autoref{sec:HomogeneousMagneticField}) via
$\rho_{\phi} = - \frac{1}{2} (\wideparen{\rho_{\phi}}_{k^+} + \wideparen{\rho_{\phi}}_{k^-})$  and $\rho_{\theta} = \wideparen{\rho_\theta}_{k^+} - \wideparen{\rho_\theta}_{k^-}$.

Note that all linearization parameters (\autoref{eq:seg_lin_rho_theta} to \autoref{eq:seg_lin_nu}) depend on known parameters given by the circle solution (see \autoref{sec:HomogeneousMagneticField}), which makes it very easy to calculate the track direction.

\paragraph{Hit position dependence:}
The track angles can be related to the pre-fit triplet parameters (denoted by superscripts $^\textrm{pre}$ or $^C$) using the fitted residuals:
\begin{align}
  \phi_{k^\pm} \, & = \, \varphi^\textrm{pre}_{k^\pm} \; - \; {\vec {\bs \delta}}^\top {\vec {\bs f}}^\pm_{{\varphi_{k}}}    \nonumber \\
                  &  \quad \mp \;  \frac{1}{2}  \left[ \nu^\textrm{pre}_{k^\pm} \,  + \, \curv \, \rho^\textrm{pre}_{\phi_{k^\pm}}  - {\vec {\bs \delta}}^\top \left( {\vec {\bs f}}^\pm_{{\nu_{k}}}   +  \curv \, {\vec {\bs f}}^\pm_{\rho_{\phi_{k}}} \right)       \right] \label{eq:track_phi_C}  \\
 \theta_{k^\pm} \, & \, = \; {\theta}_{k^\pm}^{\rm C} \; - \; {\vec {\bs \delta}}^\top {\vec {\bs f}}^\pm_{{\theta_{k}}}    \nonumber \\
                  &  \quad + \,   \left[ \tau^\textrm{pre}_{k^\pm} \,  + \, \curv \, \rho^\textrm{pre}_{\theta_{k^\pm}}  -  {\vec {\bs \delta}}^\top \left( {\vec {\bs f}}^\pm_{{\tau_{k}}}  \, + \, \curv \, {\vec {\bs f}}^\pm_{\rho_{\theta_{k}}} \right)     \right]    \label{eq:track_theta_C} ,
\end{align}
The vectors ${\vec {\bs f}}_{X}^\pm$  are Jacobians whose elements are defined for each hit by the derivatives:
\begin{align}
  ({\vec {\bs f}}^\pm_X)_{k'} \;  := \; \frac{\partial \, X_{\pm} }{\partial \, \vec \delta_{k'}} \quad
\textrm{with~~} X=\{\varphi_{k},\, \theta_{k},\, \nu_{k},\, \tau_{k},\, \rho_{\phi_{k}},\, \rho_{\theta_{k}} \}.
\end{align}

\subsubsection{Small Bending Approximation}
In the small bending limit, i.e.  $\Phi_{k\, k+1} \ll 1$, the segment length approaches the chord length.
The bending angles can then be approximated by:
\begin{align}
\Phi_{k^\pm} \;& \approx \; \pm \, \curv \; || \vec x_{k \pm 1} - \vec x_{k}  || .
\end{align}
and the track direction at hit~$k$ is then given by:
\begin{align}
  \phi_{k^\pm} \, & = \; \varphi_{k^\pm}   \, - \, {\vec {\bs \delta}}^\top {\vec {\bs f}}^\pm_{{\varphi_{k}}}    \; \mp \; \frac{\curv}{2}   \,
                    \left(    || \vec x_{k \pm 1} - \vec x_{k}|| -  {\vec {\bs \delta}}^\top   \right)  {\vec {\bs f}}^\pm_{\rho_{\phi_{k}}}  , \\
  \theta_{k^\pm} \; & = \; \theta_{k^\pm} \, - \, {\vec {\bs \delta}}^\top {\vec {\bs f}}^\pm_{{\theta_{k}}} .
\end{align}

\subsection{Track Direction Uncertainties}
The track direction uncertainties are derived by propagating both the momentum and spatial hit uncertainties  in \autoref{eq:track_phi_C} and \autoref{eq:track_theta_C}.
Note that the azimuthal and polar angle uncertainties are fully correlated due to the requirement of momentum conservation.
One  obtains:

{
\scriptsize
\begin{align}
  \Var[{\theta_{\pm}}] \; &  = \; \left[{\vec {\bs f}}_\theta^{\pm} + {\vec {\bs f}}_\tau^{\pm} + \curv \, {\vec {\bs f}}_{\rho_{\theta}}^\pm    \right]^{\!\top}  \, {\Covvvb}_{\delta} \,   \left[{\vec {\bs f}}_\theta^{\pm} + {\vec {\bs f}}_\tau^{\pm} + \curv \, {\vec {\bs f}}_{\rho_{\theta}}^\pm    \right] \nonumber  \\
                    & \hspace{5mm} - 2  \,  \left[{\vec {\bs f}}_\theta^{\pm} + {\vec {\bs f}}_\tau^{\pm} + \curv \, {\vec {\bs f}}_{\rho_{\theta}}^\pm   \right]^{\!\top} \, \Covvb_{\delta/\curv} \,  \left[{\rho_{\theta_\pm}}  - {\vec {\bs \delta}}^\top  {{\vec {\bs f}}^\pm_{\rho_{\theta}}}    \right]   \nonumber  \\
    & \hspace{5mm} +  \sigma_\curv^2 \, \left[{\rho_{\theta_\pm}}  - {\vec {\bs \delta}}^\top  {{\vec {\bs f}}^\pm_{\rho_{\theta}}}  \right]^2  , \label{eq:variance_theta}
\end{align}
\begin{align}
 \Var[{\phi_{\pm}}] \, & = \, \left[ {\vec {\bs f}}_\varphi^{\pm}  \mp  \frac{1}{2} \left( {\vec {\bs f}}_\nu^{\pm} + \curv \, {\vec {\bs f}}_{\rho_{\phi}}^\pm  \right)  \right]^{\!\top}  {\Covvvb}_{\delta}  \left[ {\vec {\bs f}}_\varphi^{\pm} \mp \left( {\vec {\bs f}}_\nu^{\pm} + \curv \, {\vec {\bs f}}_{\rho_{\phi}}^\pm  \right)  \right] \nonumber \\
                    & \hspace{5mm} \pm   \,    \left[ {\vec {\bs f}}_\varphi^{\pm}  \mp  \frac{1}{2} \left( {\vec {\bs f}}_\nu^{\pm} + \curv \, {\vec {\bs f}}_{\rho_{\phi}}^\pm  \right)  \right]^{\!\top} \,   \Covvb_{\delta/\curv} \,   \left[ {\rho_{\phi_\pm}} -  {\vec {\bs \delta}}^\top  {\vec {\bs f}}^\pm_{\rho_{\phi}}     \right]      \nonumber  \\
                               & \hspace{5mm}    + \frac{1}{4} \sigma_\curv^2 \,  \left[ {\rho_{\phi_\pm}} -  {\vec {\bs \delta}}^\top  {\vec {\bs f}}^\pm_{\rho_{\phi}}     \right]^2    \label{eq:variance_phi} .
\end{align}
}

The first term in \autoref{eq:variance_theta} and \autoref{eq:variance_phi}  comes from the hit position shifts, the last term from the curvature shifts, and the middle term is a combination of both. 

The covariance between azimuthal and polar track angle is given by:
{\scriptsize  
\begin{align}
\textrm{Cov}[{\theta_{\pm}},{\phi_{\pm}}] \, & = \, \left[{\vec {\bs f}}_\theta^{\pm} + {\vec {\bs f}}_\tau^{\pm} + \curv \, {\vec {\bs f}}_{\rho_{\theta}}^\pm    \right]^{\!\top} \, {\Covvvb}_{\delta} \, \left[ {\vec {\bs f}}_\varphi^{\pm}  \mp  \frac{1}{2} \left( {\vec {\bs f}}_\nu^{\pm} + \curv \, {\vec {\bs f}}_{\rho_{\phi}}^\pm  \right)  \right]   \nonumber \\
                    & \hspace{3mm} \pm \frac{1}{2}  \left[{\vec {\bs f}}_\theta^{\pm} + {\vec {\bs f}}_\tau^{\pm} + \curv \, {\vec {\bs f}}_{\rho_{\theta}}^\pm    \right]^{\!\top}  \, \Covvb_{\delta/\curv} \, \left( {\rho_{\phi_\pm}} -  {\vec {\bs \delta}}^\top  {\vec {\bs f}}^\pm_{\rho_{\phi}}     \right)     \nonumber  \\
                    & \hspace{3mm} - \left[ {\vec {\bs f}}_\varphi^{\pm}  \mp  \frac{1}{2} \left( {\vec {\bs f}}_\nu^{\pm} + \curv \, {\vec {\bs f}}_{\rho_{\phi}}^\pm  \right)  \right]^{\!\top}  \, \Covvb_{\delta/\curv} \,  \left({\rho_{\theta_\pm}}  - {\vec {\bs \delta}}^\top  {{\vec {\bs f}}^\pm_{\rho_{\theta}}}  \right)   \nonumber  \\
                    & \hspace{3mm} \mp \frac{1}{2} \, \sigma_\curv^2 \, \left({\rho_{\theta_\pm}}  - {\vec {\bs \delta}}^\top  {{\vec {\bs f}}^\pm_{\rho_{\theta}}}  \right)    \, \left( {\rho_{\phi_\pm}} -  {\vec {\bs \delta}}^\top  {\vec {\bs f}}^\pm_{\rho_{\phi}}     \right)      \label{eq:covariance_phitheta} .
\end{align}
}

The covariance between the track angles and the curvature is given by:
{
\scriptsize
\begin{align}
  \textrm{Cov}[\curv,{\theta_{\pm}}] \, & = \,  \sigma_\curv^2 \, \left({\rho_{\theta_\pm}}  - {\vec {\bs \delta}}^\top  {{\vec {\bs f}}^\pm_{\rho_{\theta}}}  \right)
                                          \, - \,   \left[{\vec {\bs f}}_\theta^{\pm} + {\vec {\bs f}}_\tau^{\pm} + \curv \, {\vec {\bs f}}_{\rho_{\theta}}^\pm    \right]^\top  \Covvb_{\delta/\curv}   , \\
  \textrm{Cov}[\curv,{\phi_{\pm}}] \, & = \, \mp \, \frac{1}{2} \,   \sigma_\curv^2 \, \left( {\rho_{\phi_\pm}} -  {\vec {\bs \delta}}^\top  {\vec {\bs f}}^\pm_{\rho_{\phi}}     \right)
 \nonumber \\ &  \quad  \hspace{1cm}                                       
 \, - \,  \left[ {\vec {\bs f}}_\varphi^{\pm}  \mp  \frac{1}{2} \left( {\vec {\bs f}}_\nu^{\pm} + \curv \, {\vec {\bs f}}_{\rho_{\phi}}^\pm  \right)  \right]^{\!\top}  \Covvb_{\delta/\curv}   .
\end{align}
}

Finally, the covariance between the track angles and the track position is given by:
{
\scriptsize
\begin{align}
  \textrm{Cov}[{\vec {\bs \delta}}, {\theta_{\pm}}] \, & = \, - \, {\Covvvb}_{\delta} \, \left[{\vec {\bs f}}_\theta^{\pm} + {\vec {\bs f}}_\tau^{\pm} + \curv \, {\vec {\bs f}}_{\rho_{\theta}}^\pm    \right]
   +    \Covvb_{\delta/\curv}   \, \left({\rho_{\theta_\pm}}  - {\vec {\bs \delta}}^\top  {{\vec {\bs f}}^\pm_{\rho_{\theta}}}  \right)        ,  \\
  \textrm{Cov}[{\vec {\bs \delta}}, {\phi_{\pm}}] \, & = \, - \, {\Covvvb}_{\delta} \,   \left[ {\vec {\bs f}}_\varphi^{\pm}  \mp  \frac{1}{2} \left( {\vec {\bs f}}_\nu^{\pm} + \curv \, {\vec {\bs f}}_{\rho_{\phi}}^\pm  \right)  \right]
\nonumber \\
&  \quad  \hspace{2cm}                                                        
 \, \mp \frac{1}{2}  \, {\Covvb}_{\delta/\curv} \, \left({\rho_{\phi_\pm}}  - {\vec {\bs \delta}}^\top  {{\vec {\bs f}}^\pm_{\rho_{\phi}}}  \right)    \,                                                        .
\end{align}
}

All formulas above have been validated by a numerical simulation.

\section{Gap Spectrometer Dipole} \label{sec:spectrometer_dipole}

\begin{figure}[tb]
  \begin{center}
   \begin{picture}(\textwidth,190)
     \put(-5,0){
       \epsfig{file=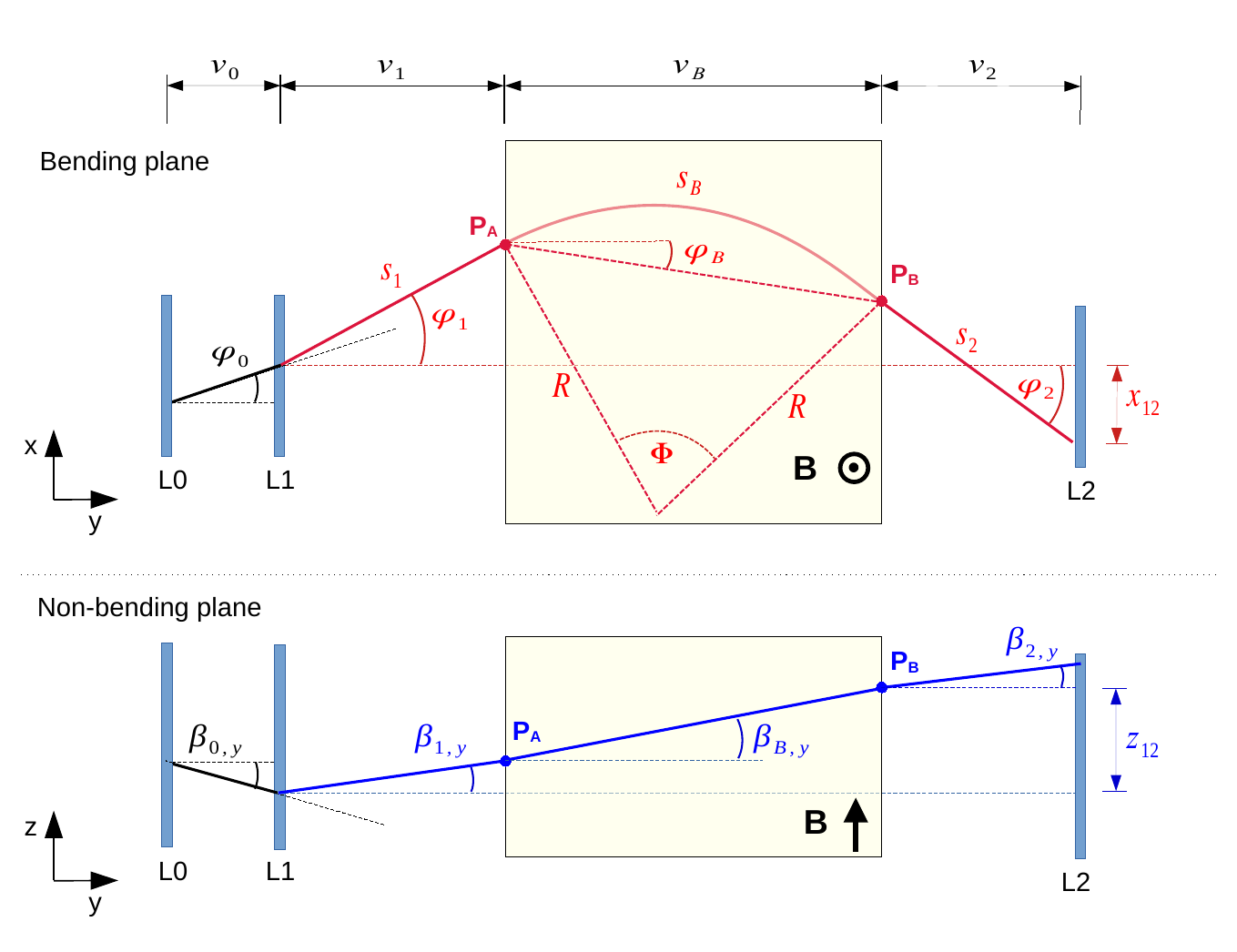,width=0.5\textwidth}
     }
   \end{picture}
   \caption{Sketch of a gap spectrometer dipole  with three layers (L0-L2) with L1 being the scattering layer of the triplet.
     The upper half shows the ($x$-$y$) bending plane and the definition of the azimuthal angles for the different track segments and the bending angle $\Phi$; the lower half shows the ($y$-$z$) plane with the elevation angles $\beta$ projected to the $y$-axis.
   }
\label{fig:sketch_dipole}
  \end{center}
\end{figure}

\autoref{fig:sketch_dipole} shows the simplest spectrometer setup with three detector planes (triplet) and one gap dipole.
Note that the notation of variables here differs from the case of a uniform magnetic field in the previous section.
The gap dipole  is placed between detector layers~1 and~2 at distances $\nu_1$ and $\nu_2$, respectively, and has a gap size of $\nu_\textrm{B}$.
In the following,
it is assumed that all measurement planes and the gap dipole entrance and exit windows are parallel.
The magnetic field is pointing in $z$-direction such that the $x$-$y$ plane is the bending plane.

The elevation angle, $\beta$, measures the particle angle with respect to the $x$-$y$ plane and is related to the polar angle via $\theta=\pi/2-\beta$ .
Note that the elevation angle is an invariant in the plane transverse to the magnetic field, i.e., $\beta_1=\beta_\textrm{B}=\beta_2$.

Since the region between tracking layers~L0 and~L1 has no magnetic field, it is easy to calculate the angles $\varphi_0$ and  $\beta_0$ for the first segment.
For the trajectory between tracking layers~L1 and~L2, the situation is far more complicated: the trajectory has first a field-free segment, then a magnetic field segment, and then another field-free segment. \\

First, the trajectory  in the bending plane is calculated.
The bending angle, $\Phi$, is defined as ratio of the arc length, $s_B$, over the bending radius, $R$.
Both parameters are related to the gap size, $\nu_\textrm{B}$, through:
\begin{align}
  \Phi &=\; \frac{s_B}{R}
\;=\; 2 \; \arcsin{\frac{\nu_\textrm{B}}{2R \cos{\varphi_\textrm{B}}}}       \label{eq:dipole_alphaRphiB} ,
\end{align}
with $\cos{\varphi_\textrm{B}}$ being the average azimuthal angle in the field region\footnote{In this calculation, stray fields are neglected.}.
The azimuthal angles in the field-free region are related to the bending angle in the field via:
\begin{align}
  {\varphi_{1,2}}  &= \;  {\varphi_\textrm{B}} \, \mp \, \frac{\Phi}{2}   .
\end{align}
With  the above relation, the bending radius in \autoref{eq:dipole_alphaRphiB} is calculated as function of the azimuthal angle~$\varphi_{1}$:
\begin{align}
R &=\;   \frac{\nu_\textrm{B}}   {\sin({\varphi_1}+\Phi)   -   \sin{\varphi_{1}}} . \label{eq:spectro_radius}
\end{align}

The horizontal offset, $x_{12} := x_2-x_1$, measured between tracking layers~L1 and~L2, is a function of  all azimuthal angles: 
\begin{align}
  x_{12}          & \; =\; \nu_1 \tan{\varphi_1}    +  \nu_\textrm{B} \tan{\varphi_\textrm{B}}  +  \nu_2 \tan{\varphi_2} \label{eq:dipole_phi1phiBphi2} \\
         & \; =\; \nu_1 \tan{\varphi_1}    +  \nu_\textrm{B} \tan{\left(\varphi_1+\frac{\Phi}{2}\right)}  +  \nu_2 \tan{\left(\varphi_1 + \Phi \right)}. \nonumber 
\end{align}
Since $\nu_1$, $\nu_2$ and $\nu_3$ are geometric constants, and  $x_{12}$ is a  measured quantity, \autoref{eq:dipole_phi1phiBphi2} defines a unique relationship between the azimuthal angle $\varphi_1$ and the bending angle $\Phi$.
This relation, however, is highly non-linear in both $\varphi_1$ and $\Phi/2$, making it difficult to derive a solution for the general case (see also \autoref{sec:spectrometer_general_solution}). \\

In the non-bending plane, the elevation angle, $\beta_1$, is given through the relation:
\begin{align}
\tan{\beta_1}  &  \; = \; \frac{z_{12}}{ s_{1}+s_B+s_{2} }  \label{eq:dipole_y12_dbd} ,
\end{align}
with  $z_{12}$ being the measured vertical offset, $  z_{12}  :=    z_2-z_1$.
The parameters $s_1$ and $s_2$ are the lengths of the no-field track segments in the bending plane,
given by:
\begin{align}
s_1 &=\; \nu_1 \, \sec {\varphi_1} , \\
s_2 &=\; \nu_2 \, \sec ({\varphi_1+\Phi_{\varphi_1}}) .
\end{align}
Here the notation $\Phi_{\varphi_1} := \Phi (\varphi_1)$ is used, where the index indicates the functional dependence given by \autoref{eq:dipole_phi1phiBphi2}.

Finally, using the bending radius (\autoref{eq:spectro_radius}) and the relation for the elevation angle (\autoref{eq:dipole_y12_dbd}) the 3D curvature is calculated as:
\begin{align}
  {\curv}  &=\; R^{-1} \, {\cos{\beta_1}} \;
             =\; \frac{\sin(\varphi_1 +\Phi_{\varphi_1}) -\sin{\varphi_1}}
    {\nu_\textrm{B}} \; \times \; \label{eq:dipole_c3d_phi_Phi}\\
  &  \frac{1}{\sqrt{
                 1+\frac{z_{12}^2}
                 {
                \left( \nu_1 \sin{\varphi_1} \, + \,  \nu_2 \sin(\varphi_1+\Phi_{\varphi_1}) \,  + \, 
                \nu_\textrm{B} \frac{\Phi_{\varphi_1} }{\sin(\varphi_1 +\Phi_{\varphi_1}) -\sin{\varphi_1}}
                \right)^2                 
                 }
                 }} .   \nonumber 
\end{align}
This equation is  non-linear in $\varphi_1$ and $\Phi_{\varphi_1}$.
Since the function $\Phi({\varphi_1})$ is also non-linear,
the linearization around a reference solution is here much more complicated than in the case of a triplet in a uniform magnetic field (\autoref{sec:HomogeneousMagneticField}).

As motivated in \autoref{sec:method}, the solution with zero kink angle in the bending plane, $\varphi_{1,\textrm{ref}} =\ \varphi_{0}$, is chosen as reference for the linearization.
In the following, the solution for the general case and the small bending approximation is discussed.

\subsection{General Solution} \label{sec:spectrometer_general_solution}

The general solution involves  solving \autoref{eq:dipole_phi1phiBphi2} numerically to determine the reference bending angle $\Phi_\textrm{ref} = \Phi(\varphi_{1,\textrm{ref}})$.
With the help of $\Phi_\textrm{ref}$, all parameters of the reference trajectory can be calculated;
most importantly, the reference elevation angle $\beta_\textrm{ref}$ (\autoref{eq:dipole_y12_dbd}), and the reference curvature $\curvref$ (\autoref{eq:dipole_c3d_phi_Phi}). 

For the determination of the $\rho$ parameters, the derivatives $\diff{\curv}/\diff{\varphi_{1}}$ and
$\diff{\curv}/\diff{\theta_{1}}$ are needed.
An analytical solution can be derived by differentiation of \autoref{eq:dipole_c3d_phi_Phi} and \autoref {eq:dipole_y12_dbd}, respectively, and by using the relations:
\begin{align}
  \rho_\theta \;  &  =\; \rho_\phi \, \frac{\partial{(\Delta \theta)}}{\partial{(\Delta \phi})} 
              \;  =\; \rho_\phi \, \frac{\partial{\theta_{1}}}{\partial{\varphi_{1}}} 
              \;  =\; - \rho_\phi \, \frac{\partial{\beta_{1}}}{\partial{\varphi_{1}}} .
\end{align}
However, the equations obtained in this way will be very  unwieldy.
In fact, it is much easier to determine the $\rho$ parameters numerically from a small variation, $\epsilon$, of the azimuthal angle $\varphi_{1,\textrm{ref}} \rightarrow \varphi_{1,\textrm{ref}}^{\, \epsilon} = \varphi_{1,\textrm{ref}} +\epsilon_\varphi$.
This yields a second solution, which is denoted as $\curvref^{\, \epsilon}$ and $\beta_{1,\textrm{ref}}^{\, \epsilon}$.
The fundamental triplet parameters are then given by:
\begin{align}
  \rho_\theta &=\; - \frac{\beta_{1,\textrm{ref}}^{\, \epsilon}-\beta_{1,\textrm{ref}}}
  {\curvref^{\, \epsilon} - \curvref} , \\
  \rho_\phi &=\; \frac{\epsilon_\varphi}
              {\curvref^{\, \epsilon} - \curvref} , \\
  \tilde \Theta &=\; - \rho_\theta \, \curvref \,-\,  (\beta_{1,\textrm{ref}} - \beta_0) , \\
  \tilde \Phi &=\; - \rho_\phi \, \curvref .
\end{align}

\subsection{Approximation for Small Bending Angles}
In the case of small bending angles $\Phi$, \autoref{eq:dipole_phi1phiBphi2} can be linearized  to first order in $\Phi$.
In this approximation, the fundamental triplet parameters are given by:
\begin{align}
  \tilde \Phi &  \approx \; \arctan \left( \frac{x_{12}}{y_{12}} \right) - \varphi_0 ,  \\
  \tilde \Theta &  \approx \; \arctan \left( \frac{z_{12}}{d_{12}} \right) - \theta_0 , \\
  \rho_\phi &  \approx \; - \frac{\sqrt{d_{12}^2+z_{12}^2}}{2} \; \frac{\nu_B}{y_{12}} \; \left( 1 - \frac{\nu_{2}-\nu_{1}}{y_{12}}   \right) , \\
  \rho_\theta & \approx \; 0 ,
\end{align}
with $d_{12}:=\sqrt{x_{12}^2+y_{12}^2}$ being the distance between hit~1 and~2 in the bending plane.
Similar to the weak bending case in a uniform magnetic field, the parameter $\rho_\theta$ vanishes here.
For symmetric setups, i.e., $\nu_{1}=\nu_{2}$,  the parameter  $\rho_\phi$ is equivalent to half the effective track length in the dipole field.
In the limit $\nu_{1}\rightarrow 0$ and $\nu_{2}\rightarrow 0$ (the tracking layer~L1 and~L2 are positioned at the dipole entrances and exit), this parameter becomes: $\rho_\phi = - {\sqrt{d_{12}^2+z_{12}^2}}/{2}$.
Similar to the uniform magnetic field setup, the $\rho_\phi$ parameter corresponds to half the length that the particle travels in the magnetic field, consistent with the naive expectation from the Lorentz force law. \\

\section{Triplet with General Field Configuration} \label{sec:GeneralFieldConfiguration}

An inhomogeneous or irregular magnetic field represents the greatest challenge for the calculation of the triplet parameters.
The main difficulty is that trajectories connecting the hit positions $\vec x_{0}$,  $\vec x_{1}$, and  $\vec x_{2}$ need to be found, and this by means of track extrapolation since no analytical solution exists, see also \autoref{fig:triplet_approximation}.

In the following, an algorithm is presented that finds a reference trajectory for an arbitrary (inhomogeneous) magnetic field by means of track extrapolation.
This algorithm will be used to calculate the fundamental triplet parameters and the hit gradients.

\subsection{Finding a Reference Trajectory} \label{sec:GeneralFieldReferenceTrajectory}
In the following, it is assumed that an approximate solution for the hit triplet exists and is known.
This could be, for example, a solution obtained with a constant averaged magnetic field strength.
The reference trajectory can then be found using Newton's method in 2~dimensions ($\theta-\phi$ space).
A procedure tailored for hit triplets is described in the following paragraphs.

\begin{figure}[tb]
   \begin{picture}(0.5\textwidth,140)
     \put(-5,0){
       \epsfig{file=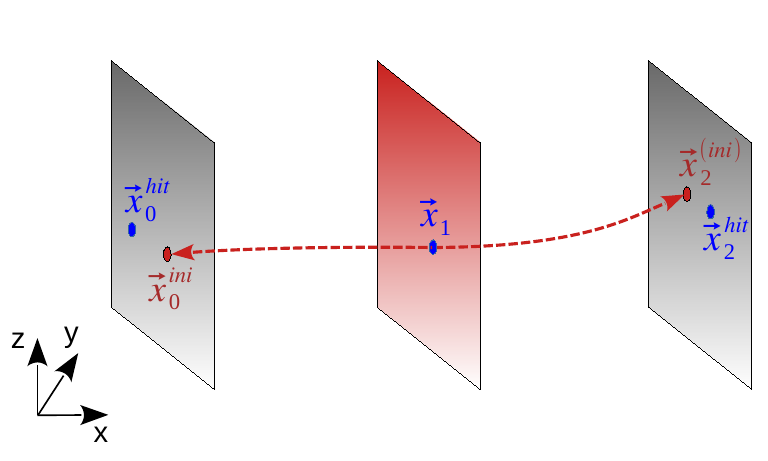,width=0.49\textwidth}
     }
   \end{picture}
   \caption{Sketch showing the first step of the extrapolation algorithm for inhomogeneous magnetic fields. The \textsl{red dashed} line  shows the trajectory obtained by extrapolating the track vector at the middle layer
  from the hit position ${\vec x}_1$ to both sides. }
\label{fig:triplet_approximation}
\end{figure}

At the middle hit position, the initial approximated solution is described by five track parameters:
\begin{itemize}
\item $\curv^\textrm{ini}$: 3D curvature of the approximate trajectory;
\item $\theta_{10}^\textrm{ini}$ ($\theta^\textrm{ini}_{12}$): polar angle of the track at the middle hit position before (after) scattering;
\item $\phi^\textrm{ini}_{10}$ ($\phi^\textrm{ini}_{12}$): azimuthal angle of the track at the middle hit position before (after) scattering.
\end{itemize}

By extrapolating the trajectory from the middle hit position to both sides, as sketched in \autoref{fig:triplet_approximation}, the extrapolated hit positions  $\vec x_{0}^\textrm{\,ini}$ and  $\vec x_{2}(0)=\vec x_{2}^\textrm{\,ini}$ are obtained.
In the next step, the polar and azimuthal angles are varied by  $\epsilon_\theta$ and $\epsilon_\phi$, respectively,
and a new set of angles is obtained for track extrapolations:
\begin{align}
  \renewcommand\arraystretch{1.5}
\left.
  \begin{array}{cc}
\theta_{1k}^{(\epsilon_\theta)}  &=\; \theta_{1k}^\textrm{ini} +  \epsilon_\theta \\
\phi_{1k}^{(\epsilon_\phi)}  &=\; \phi_{1k}^\textrm{ini} +  \epsilon_\phi 
  \end{array}
 \quad  \right\} \quad \ k=0,2
\end{align}
Two more extrapolations  to the tracking planes L0 and L2 are performed on each side.
Let  $\vec x_{k}^{\,(\epsilon_\theta)}$ and $\vec x_{k}^{\,(\epsilon_\phi)}$ be the  extrapolated points from the  polar angle and azimuthal angle variations, respectively,
the matching condition reads:
\begin{align}
 \vec x_k^{\,\prime} & = \, \label{eq:GeneralFieldMatchingCondition} 
             \; \vec x_k \, + \, \eta_{\theta,k} \; \frac{\vec x_k^{\,(\epsilon_\theta)} - \vec x_k^\textrm{\,ini}}{\epsilon_\theta} 
                          \,  + \, \eta_{\phi,k} \; \frac{\vec x_k^{\,(\epsilon_\phi)} - \vec x_k^\textrm{\,ini}}{\epsilon_\phi}
 \; \eqex \;  \vec x_k^\textrm{\, hit} 
                                                            , \nonumber
\end{align}
with $\vec x_k^{\,\prime}$ being the calculated interception points and $\vec x_k^\textrm{\, hit}$ being the measured hit positions.
For each side ($k=0,2$), a system of three equations with two unknown correction parameters:  $\eta_{\theta,k}$ and $\eta_{\phi,k}$ are obtained.
The third degree of freedom corresponds to the segment length, which is indirectly determined by the extrapolation procedure.

Above system of equations can be solved by minimizing the spatial distance $|| \vec x_k^\textrm{\,hit} - \vec x_k^{\,\prime} ||$.
Using the short-hand notations:
\begin{align}
\vec a_{\theta,k} \;&:=\; \vec x_k^{\,(\epsilon_\theta)} - \vec x_k^\textrm{\,ini}  \nonumber , \\
\vec a_{\phi,k} \;&:=\; \vec x_k^{\,(\epsilon_\phi)} - \vec x_k^\textrm{\,ini} \nonumber ,
\end{align}
the solution for each side ($k=0,2$) is given by:
\begin{align}
  \left(
\begin{array}{c}
  \eta_{\theta,k} / \epsilon_\theta \\
  \eta_{\phi,k}  / \epsilon_\phi
\end{array}
  \right)
  \;&=\;
  \;  \frac{\vec x_k^\textrm{\,hit} - \vec x_k^\textrm{\,ini}}
  {\vec a_{\theta,k}^{\,2} \, \vec a_{\phi,k}^{\,2} - (\vec a_{\theta,k} \, \vec a_{\phi,k})^2 }
      \\    &  \hspace{0.3cm}
  \cdot \  \left(
\begin{array}{c}
    \vec a_{\phi,k}^{\,2} \; \vec a_{\theta,k} - (\vec a_{\theta,k} \; \vec a_{\phi,k}) \; \vec a_{\phi,k} \\
  \vec a_{\theta,k}^{\,2} \; \vec a_{\phi,k} - (\vec a_{\theta,k} \; \vec a_{\phi,k}) \; \vec a_{\theta,k}
\end{array} 
              \right)
\begin{array}{c}
           \\   .
\end{array} 
              \nonumber
\end{align}

In case of non-linearities, the matching condition might not be fulfilled in one extrapolation step and the procedure needs to be iterated until $\eta_{\theta,k}$ and $\eta_{\phi,k}$ are determined with the required accuracy, i.e., $|| \vec x_k^{\, \prime} - \vec x_k^\textrm{\,hit}  || < \textrm{accuracy}$.
Finally, the track parameters for the reference solution at the middle layer are given by:
\begin{align}
 \theta_{1k} &=\; \theta^\textrm{ini}_{1k} + \eta_{\theta_{k}} , \\
 \phi_{1k} &=\; \phi^\textrm{ini}_{1k} + \eta_{\phi_{k}}  .
\end{align}

\subsection{Determination of the Triplet Parameters} \label{sec:GeneralFieldTripletParameters}
For the determination of the $\rho$ parameters, the curvature is varied according to:
\begin{align}
\curv^{(\epsilon_\curv)}  &=\; \curv^\textrm{ini} +  \epsilon_{\curv} .
\end{align}
Using the so modified curvature, the trajectory is extrapolated again from the middle layer to both sides.
Similar to the procedure described in \autoref{sec:GeneralFieldReferenceTrajectory},
a new set of polar $\theta_{1k}^{\,(\epsilon_{\curv})}$  and azimuthal $\phi_{1k}^{\,(\epsilon_{\curv})}$ angles is determined such that both reconstructed track segments match the actual hit positions.
The four fundamental triplet parameters are then obtained from:
\begin{align}
  \rho_\theta &=\; \frac{\theta_{12}^{\,(\epsilon_{\curv})}  - \theta_{10}^{\,(\epsilon_{\curv})} - {\theta_{12}} + {\theta_{10}}}
  { \epsilon_{\curv}} , \\
  \rho_\phi &=\; \frac{\phi_{12}^{\,(\epsilon_{\curv})}  - \phi_{10}^{\,(\epsilon_{\curv})} - \phi_{12} + \phi_{10}}
  { \epsilon_{\curv}} , \\
  \tilde \Theta &=\; - \rho_\theta \, \curv^\textrm{ini} \,+\,  \theta_{12} - \theta_{10}   , \\
  \tilde \Phi &=\; - \rho_\phi \, \curv^\textrm{ini}  \,+\,  \phi_{12} - \phi_{10} .
\end{align}

\subsection{Determination of the Hit Gradients} \label{sec:GeneralFieldHitGradients}
For the determination of the hit gradients (see \autoref{sec:Representation}), no further track extrapolations are required.
By solving \autoref{eq:GeneralFieldMatchingCondition} for the in total $3 \times 3$ $1\sigma$ hit position shifts, a set of nine polar and azimuthal angle shifts is obtained, which serves as input for the calculation of the hit gradients according to \autoref{eq:hitGradientTheta} and~\ref{eq:hitGradientPhi}. \\

To summarize, for the determination of the triplet parameters, at least four track extrapolations are needed for each segment of the triplet:
one for the starting trajectory (estimate), two for the track direction variation, and one  for the curvature variation.
In case of large non-linearities, more extrapolations might be required.
This method is universal and can be used for any tracking detector with any arbitrary field configuration.



\bibliographystyle{elsarticle-num}
\bibliography{references}

\end{document}